\documentclass[12pt]{article}

\usepackage{scicite}

\usepackage{times}
\usepackage{graphicx}
\usepackage{booktabs}
\usepackage{lineno}

\newenvironment{sciabstract}{%
\begin{quote} \bf}
{\end{quote}}

\title{Propulsion Contribution from Individual Filament in a Flagellar Bundle}

\author
{Jin Zhu,$^{1\dag}$ Yateng Qiao,$^{1\dag}$ Lingchun Yan,$^{1}$ Yan Zeng,$^{1}$ Yibo Wu,$^{1}$ \\ Hongyi Bian,$^{1}$ Yidi Huang,$^{1}$ Yuxin Ye,$^{1}$ Yingyue Huang,$^{1}$ Russell Hii Ching Wei,$^{1}$\\ Yinuo Teng,$^{1}$ Yunlong Guo,$^{1}$ Gaojin Li,$^{2\ast}$ Zijie Qu$^{1\ast}$\\
\\
\normalsize{$^{1}$UM-SJTU Joint Institute, Shanghai Jiao Tong University,}\\
\normalsize{Shanghai 200240, PR China.}\\
\normalsize{$^{2}$State Key Laboratory of Ocean Engineering, School of Ocean and Civil Engineering,}\\
\normalsize{Shanghai Jiao Tong University}\\
\normalsize{Shanghai 200240, PR China.}\\
\\
\normalsize{$^\ast$Email address for correspondence:  gaojinli@sjtu.edu.cn; zijie.qu@sjtu.edu.cn} \\
\normalsize{$^\dag$These authors contributed equally to this work.}
}

\date{}

\begin{document} 


\baselineskip24pt


\maketitle


\begin{sciabstract}
Flagellated microorganisms overcome the low-\emph{Reynolds}-number time reversibility by rotating helical flagella \cite{purcell_life_1977, bray_cell_2001, lauga_hydrodynamics_2009, lauga_bacterial_2016}. For peritrichous bacteria, such as \textit{Escherichia coli}, the randomly distributed flagellar filaments align along the same direction to form a bundle, facilitating complex locomotive strategies \cite{berg_chemotaxis_1972, turner_real-time_2000, Darnton_Turner}. To understand the process of flagella bundling, especially the propulsion force, we develop a multi-functional macroscopic experimental system and employ advanced numerical simulations for verification. Flagella arrangements and phase differences between helices are investigated, revealing the variation in propulsion contribution from the individual helix. Numerically, we build a time-dependent model to match the bundling process and study the influence of hydrodynamic interactions. Surprisingly, it is found that the total propulsion generated by a bundle of two filaments is constant at various phase differences between the helices. However, the difference between the propulsion from each helix is significantly affected by the phase difference, and only one of the helices is responsible for the total propulsion at a phase difference equals to $\pi$. Through our experimental and computational results, we provide a new model considering the propulsion contribution of each filament to better understand microbial locomotion mechanisms, especially on the wobbling behavior of the cell. Our work also sheds light on the design and control of artificial microswimmers.

\end{sciabstract}


\section*{Introduction}
Microorganisms navigate through complex aquatic environments by beating flexible cilia or rotating rigid helical flagella \cite{bray_cell_2001}. Understanding the swimming strategies of various cells at low-\emph{Reynolds}-number environments is crucial for revealing underlying biological processes such as surface colonization \cite{ottemann_helicobacter_2002}, tissue invasion \cite{horstmann_flagellin_2017}, and escape from harmful substances \cite{valentine_propane_2010}. This knowledge is also pivotal in designing artificial microswimmers with potential engineering applications \cite{zhang_characterizing_2009,ahmed_artificial_2016,yan_multifunctional_2017,choi_recent_2021,deng_acoustically_2023}.

Over the past few decades, the swimming behavior of \textit{Escherichia coli}, a typical peritrichous bacterium, has been extensively studied. Berg and Brown were pioneers in using a 3D real-time tracking microscope to study the motion of wild-type \textit{E. coli}, characterized by a ``run-and-tumble'' behavior \cite{berg_chemotaxis_1972}. During the ``run'' phase, \textit{E. coli} cells align their randomly distributed flagellar filaments in the same direction to form a bundle. They then unbundle and change direction during the ``tumble'' phase by rotating their motor(s) in the opposite direction \cite{turner_real-time_2000,Darnton_Turner,lauga_bacterial_2016}.

Numerous studies have been conducted to understand the complex interactions of flagella from various perspectives. For instance, Turner \textit{et al.} \cite{turner_real-time_2000} and Darnton \textit{et al.} \cite{Darnton_Turner} employed fluorescently labeled bacteria to visually demonstrate the flagella bundling process. Macnab \cite{Macnab1977} and Kim \textit{et al.} \cite{KIM2003} conducted macroscopic experiments using scaled models to show that the bundling process is a mechanical phenomenon induced by hydrodynamic interactions. Recently, Lim \textit{et al.} utilized a centimeter-scale multi-flagellated robot to further explore the bundling sequence \cite{lim_bacteria-inspired_2023}. Additionally, many numerical studies have been performed \cite{TONG2023101924, lee2018bacterial}, including those considering the polymorphic transformations of flagella \cite{turner_real-time_2000}. Lim and Peskin \cite{PhysRevE.85.036307} applied the immersed boundary method, indicating that bundling occurs when both flagella are left-handed helices turning counterclockwise (viewed from the non-motor end looking back toward the motor) or when both are right-handed helices turning clockwise. Lee \emph{et al.} \cite{lee2018bacterial} employed Kirchhoff rod theory to analyze the polymorphic transformations during the flagella bundling and unbundling process.

Historically, most experimental and theoretical models have treated the flagella bundle as a single helix \cite{Hancock_1953,HOLWILL1963249,keller_swimming_1976,magariyama_mathematical_2002,lauga_hydrodynamics_2009,patteson_running_2015,lauga_bacterial_2016}. However, recent studies on motor torque re-allocation among flagella \cite{kamdar_multiflagellarity_2023} and synchronization of flagella \cite{reichert_synchronization_2005,qian_minimal_2009, reigh_synchronization_2012,tatulea-codrean_elastohydrodynamic_2022} reveal significant differences in the contribution of each helix during the bundling process. Additionally, little research has focused on propulsion during the ``tumble'' phase or the flagellar bundling process itself. Overall, a comprehensive study of propulsion generation from individual flagellar filaments and the total propulsion caused by a formed bundle is still lacking.

In this work, we extend the research of Kim \textit{et al.} \cite{KIM2003} by conducting an experimental study using a scaled model. We directly measure the propulsive force during the bundling process of two helices rotating at constant speeds. Our results indicate that total propulsion decreases as the helices bundle, primarily due to hydrodynamic interactions. These findings are corroborated by a computational framework where a solid-fluid coupled system is employed. Additionally, we observe that while the phase difference between the helices does not alter the total propulsion, it significantly affects the force distribution, with one helix generating almost all the propulsion when the phase difference is $\pi$. This outcome provides a new explanation for cell precession (wobble phenomenon) and is further supported by simplified 2D theoretical work. Contrary to several existing studies \cite{patteson_running_2015,PhysRevFluidsQu,kamdar_colloidal_2022}, our results suggest a positive correlation between the swimming speed of the cell and the wobbling effect. We also propose that our findings can inform control strategies for artificial helical swimmers; by adjusting the phase difference between two helices, a perpendicular side force can be generated relative to the swimming direction, which could help navigate complex environments.

\section*{Results and Discussion}
The impact of the separation distance between the two helices, denoted as $c$, and the phase difference ($\Delta \Phi = \Phi_1 - \Phi_2$) on the bundling dynamics and propulsive force generation is thoroughly investigated. All the experiments are done in the multi-functional macroscopic experimental system as Fig. 1A shows (details see SI section 1.1). It is important to note that $\Delta \Phi$ remains constant throughout each experiment because the motor operates at a constant speed, mirroring the conditions found in \emph{E. coli} cells \cite{Berry_motor, turner_real-time_2000, sowa_bacterial_2008}.

\subsection*{Propulsion during the bundling process}
Firstly, simple bundling experiments are conducted. The process, documented in SI movie 1, aligns with previous studies \cite{KIM2003,kim_particle_2004}. In these experiments, we monitor the propulsion force exerted by each helix during the bundling process, with results shown in Fig. 1C\&D. The time variable ($t$) is scaled as $t^{*} = \omega t / 2\pi$, and all propulsion forces ($F$) are scaled as $F^* = F/\mu \omega R L$ in this work. The sensor measures force solely along the axial ($z$) direction as Fig. 1B shows.

An overall decaying trend in the force generated by each helix over time ($F_{1}^* (t)$ and $F_{2}^* (t)$) is observed, starting with an initial value of $F_{single}^* \approx 0.55$. This is consistent with predictions from traditional Resistive Force Theory (RFT) \cite{Hancock_1953,rodenborn_propulsion_2013} for a single immersed helix at low \emph{Reynolds} number (see SI section 2.1.1) and the single helix propulsion force experiment (see SI section 2.1.3). In Fig. 1C\&D, both cases show a similar overlap between the force-time curves from each helix, which indicate system symmetry. If we check the total propulsion force $F^*_{t} (t)$, it is clear that the force plateaus at $t^{*}_c \approx 140$, when the two helices contact each other and cease bundling (also detailed in SI section 2.1.4). We thus categorize the experiment into two phases: a bundling state ($t^{*} < t^{*}_c$) and a steady state ($t^{*} > t^{*}_c$).  In Fig. 1C\&D, the total propulsion, $F^*_{t} (t)$, represented by a yellow curve, is initially twice the value of propulsion generated by a single helix, indicating negligible hydrodynamic interaction at this early stage. As time progresses, $F^*_{t} (t)$ changes due to helix tilt and increase hydrodynamic interaction. At the steady state, the fluctuation of $F^*_{t} (t)$ is considerably smaller than that of $F^*_{1} (t)$ and $F^*_{2} (t)$, which is attributable to helix collisions. However, Fig. 1D shows varying $F_{1}^* (t)$ and $F_{2}^* (t)$ though the system is considered as steady state with constant $F^*_{t} (t)$ in the case of $\Delta \Phi = \pi$ rad. The consistent phenomenon for $F^*_{t} (t)$ and varying phenomenon $F_{i}^* (t)$ for individual helix are also considered as hydrodynamic interaction in later discussion.

Further examination of the cause of propulsion change focuses on the inclination of each helix at the steady state. As Fig. 1B shows, the inclination angle, $\delta_{i}$ (where $i$ = 1 and 2 for each helix), is defined as the small angle between the $z$-axis and the helix at the steady state. Propulsion excluding hydrodynamic interaction is defined as $F^*_{i, z} = F^*_{single}\cos(\delta_{i})$, where the preconditioned single helix propulsion force ($F^*_{single}$) is obtained as SI section 2.1.3. The difference between $F^*_{i}$ and $F^*_{i, z}$, mainly attributable to hydrodynamic interaction ($F^*_{i, hydro} = F^*_{i} - F^*_{i, z}$). For instance, the case of $\Delta \Phi = 0$ rad shows that $F^*_{i, z}$ = $0.48$ and $F^*_{i, hydro}$ is about $-0.12$. The propulsion reduction due to system tilt, calculated as $F^*_{single} - F^*_{i, z}$ = $-0.02$, is significantly less impactful than the hydrodynamic interaction. Also, we can find, the case of  $\Delta \Phi = \pi$ rad breaks the symmetry of the bundling system, indicating greater influence of hydrodynamic interaction since two individual helices show large force difference in the steady state. This indicates that the primary factor in the propulsion variation during the bundling process is hydrodynamic interaction.

\subsection*{The influence of helix separation and phase difference}
Peritrichous bacteria such as \emph{E. coli} have multiple flagella projecting in various directions \cite{turner_real-time_2000,BergE_coliinMotion,lauga_bacterial_2016}. Due to the cell body shape and the random anchoring of motors \cite{BergE_coliinMotion,lauga_bacterial_2016}, flagella separation varies significantly. Understanding the bundling dynamics at different separations is crucial. We adjust the separation $c^{*}$ = $c/L$ from 0.22 to 0.56.

Firstly, the measured bundling time ($T_{b}$) is investigated (see SI fig. S12A). $T_{b}^{*}$ increases non-linearly with $c^*$, which partly explains the significant variation in bundling times reported in previous studies \cite{berg_chemotaxis_1972,patteson_running_2015,qu_changes_2018}. Additionally, the total propulsion $\Sigma F^{*}$ at steady state is investigated (see Fig. 2A). Interestingly, $\Sigma F^{*}$ remains constant regardless of $c^{*}$, suggesting that for a fixed number of flagella filaments (2 in this case), the propulsion is a constant. Hence, the swimming speed of a cell primarily depends on the drag induced by the cell body shape. Moreover, due to the symmetry in these experiments, the inclination angles ($\delta_{i}$) of each helix are identical at the steady state (see SI fig. S12B).

Furthermore, we studied the effect of phase difference ($\Delta \Phi$) on propulsion. In Fig. 2B, $\Sigma F^{*}$ is plotted over various $\Delta \Phi$ values ranging from -$\pi$ to $\pi$ with an interval of $\pi/2$. While $\Sigma F^{*}$ remains constant across different $\Delta \Phi$, the difference in propulsion, $\Delta F^{*} = F_{1}^{*} - F_{2}^{*}$, generated by each helix is linearly correlated with $\Delta \Phi$. Remarkably, at $\Delta \Phi = -\pi$, $\Delta F^{*}$ approximates $\Sigma F^{*}$, indicating that the total propulsion is contributed by only one filament. This finding holds for $\Delta \Phi = \pi$ as well, by comparing -$ \Delta F^{*}$ with $\Sigma F^{*}$ due to symmetry.

What causes the propulsion difference? It is noted that with varying $\Delta \Phi$, the system loses symmetry, causing the bundle to lean toward the helix with a lagging phase at the steady state, as shown in Fig. 2C. The leaning angle, $\theta$, is plotted over different $\Delta \Phi$ values. Consequently, the inclination angle, $\delta_{i}$, is no longer identical for the helices at the steady state (see SI Table. S2). The propulsion excluding hydrodynamic interaction, $F^*_{iz}$, is calculated and plotted over $\Delta \Phi$ in Fig. 2D. However, the propulsion difference calculated without considering the hydrodynamic interaction is significantly smaller than that measured directly from the sensor (Fig. 2D, square markers). Thus, the changed hydrodynamic interaction is identified as the primary cause of the propulsion difference between each helix, with this effect increasing with $\Delta \Phi$.

\subsection*{Hydrodynamic interaction revealed by a numerical simulation}
To further elucidate the hydrodynamic interaction, we conducted a numerical simulation using a solid-fluid coupled solver implemented in OpenFOAM (see SI section 2.2). The accuracy of the numerical solver is validated by simulating the motion of a Jeffery Orbit \cite{Jeffery1922TheMO} (SI section 2.2.4) and the propulsion generated by a single helix (SI section 2.2.5), with results consistent with theoretical predictions \cite{KIM1991107}. In the simulations, each helix is modeled as a rigid body rotating around a fixed point at a constant speed, with boundary conditions, fluid properties, and initial conditions replicating those of the experimental setup.

The bundling process observed in the simulation aligns with experimental observations (Fig. 3A), further confirming the simulation's accuracy. Additionally, the propulsion values calculated from the simulation (Fig. 3B) closely match the experimental results. To visualize the interaction directly, we plotted the flow field from the simulation at various time points during the bundling process in Fig. 4A-C. Initially, the fluid velocity is nearly zero but increases rapidly as the simulation progresses. The velocity between the helices also intensifies as they bundle closer together, indicating stronger hydrodynamic interactions. The radial decay of the flow velocity $U^*$ in different time indicates the different bundling stages (Fig. 4D). The green, orange and purple lines represent the beginning of the bundling process. The blue line represent the separation process in y-direction. The red and black dash lines represent the flagella become close to each other in y-direction and. With time goes on, this velocity radial decay line indicates that the hydrodynamic interaction becomes stronger. The correlation coefficient of velocity is also shown in SI section 2.5 (see SI fig. S19). Moreover, the tendency of the helices to lean toward one during the steady state is attributed to a hydrodynamic force acting along the $x$-axis. This phenomenon is well-demonstrated in the simulation results (SI section 2.4), where the side force acting on the system is shown for cases when the phases are either $\Delta \Phi=0$ (see SI fig. S18A) or $\Delta \Phi= \pi$ (see SI fig. S18B).

\subsection*{Unbalanced propulsion leads to bacteria wobbling}
Previous studies have observed a "wobbling" trajectory in flagellated bacteria swimming, attributed to the misalignment between the bacterium's body axis and the bundle axis \cite{keller_swimming_1976,The_wiggling_trajectories_of_bacteria2012, Darnton_Turner,yin_escaping_2022}. However, the relationship between the wobbling effect and the swimming speed remains unclear \cite{Darnton_Turner,binliu_wobble,patteson_running_2015,PhysRevFluidsQu,kamdar_colloidal_2022}. In our study, the varied leaning angle, $\theta$, resulting from different phase differences, $\Delta \Phi$, also suggests potential misalignment. But is misalignment the sole reason for the wobbling behavior? To explore this, we conducted a simple theoretical analysis by modeling the cell body as a 2D ellipse defined by the equation $x^2/(a/2)^2 + y^2/(b/2)^2 = 1$, where $a = 2.5$ $\mu$m and $b = 1$ $\mu$m represent the major and minor axis lengths, respectively. The uneven propulsion is modeled as two pairs of point forces ($F_1^{\parallel}$, $F_1^{\perp}$ and $F_2^{\parallel}$, $F_2^{\perp}$) acting on the ellipse, illustrated in SI fig. S20. The angles $\delta_1$ and $\delta_2$ are defined similarly but in a 2D plane, with $F_i^{\perp} = F_i^{\parallel}\tan\delta_{i}$.

Building on previous research \cite{The_wiggling_trajectories_of_bacteria2012,patteson_running_2015}, it is suggested that the wobbling frequency is equal to the self-rotating frequency of the cell body, implying that the relative angle of the bundle axis to the body axis remains constant during the "run" phase. The motion of the bacteria is characteristic of axial precession. Given the force-free nature of micro-swimmers in low \textit{Reynolds} number environments \cite{purcell_life_1977,lauga_bacterial_2016}, a force balance analysis is essential. This analysis focuses on the angle $\beta$, which equals half of the wobble angle and is represented in the swimming direction relative to the body's major axis (see SI fig. S20) \cite{Darnton_Turner}.

\subsection*{Hydrodynamic interaction and wobbling dynamics}
The relative fluid velocity with respect to the elliptical cell body, $v_{fluid}$ ($v_{fluid} = -v$), is decomposed into two components: $v_{\parallel} = -V \cos\beta \hat{i}$ along the major axis and $v_{\perp} = -V \sin\beta \hat{j}$ along the minor axis, where $\hat{i}$ and $\hat{j}$ represent the unit vectors in the directions parallel and perpendicular to the major and minor axes, respectively. The drag coefficients for these axes, $K_{\parallel}$ and $K_{\perp}$, are derived from the geometric anisotropy and orientation of the flow as detailed by Happel and Brenner \cite{happel_low_1983} in equations 4-25.25 and 4-26.38 respectively.

\begin{equation}
    K_{\parallel} = \frac{4}{3\sqrt{\tau_0^2-1}[(\tau_0^2+1) \coth^{-1}\tau_0-\tau_0]}
\end{equation}
\begin{equation}
    K_{\perp} = \frac{4}{3\sqrt{\lambda_0^2+1}[\lambda_0 - (\lambda_0^2 -1) \cot^{-1}\lambda_0]}
\end{equation}
where $\lambda_0 = [(a/b)^2-1]^{-1/2}$ and $\tau_0 = [1-(b/a)^2]^{-1/2}$ are geometric factors.

The force equilibrium in the x-direction and y-direction are given by:
\begin{equation}
   F_{1}^{\parallel} + F_{2}^{\parallel}= F_{drag}^{\parallel}=-6\pi\eta b K_{\parallel} v_{\parallel}
\end{equation}
\begin{equation}
    F_{1}^{\perp} + F_{2}^{\perp} = F_{drag}^{\perp}=-6\pi\eta a K_{\perp} v_{\perp}
\end{equation}
where $\eta = 10^{-3}$ Pa$\cdot$s is the viscosity of the fluid. The wobble angle $2\beta$ is subsequently calculated and the results are summarized in Table 1. The ratio $\Lambda = F_2^{\parallel}/F_1^{\parallel}$ is used to align theoretical values with experimental observations. In calculations, we take the propulsion force in the x-direction ($F_{1}^{\parallel} + F_{2}^{\parallel} = 0.24$ pN) from the "no-wobble" case where $\Lambda = 1$. The results corroborate the values measured in previous studies \cite{patteson_running_2015, Darnton_Turner, yin_escaping_2022} and intriguingly, we observe a positive correlation between the swimming speed $v$ and the wobble angle, supporting the findings by Liu \textit{et al.} \cite{binliu_wobble}. Contrary to their explanation of additional propulsion generated by cell body rotation due to an asymmetric body shape, our findings suggest an enhancement in swimming speed even with an axisymmetric cell body, primarily due to the significant force imbalance between each helix which not only increases the wobbling effect but also contributes to higher propulsion and faster swimming.

\section*{Conclusion}
In this study, a macroscopic experimental system was employed to investigate the bundling dynamics of two helices rotating at constant speeds, with a focus on the changes in propulsion during the bundling process. We discovered that the phase difference between the helices leads to significant deviations in propulsion at the steady state, while the total propulsion remains unchanged. Our experimental findings, supported by numerical simulations, highlight the crucial role of hydrodynamic interactions in both the bundling process and propulsion generation.

Additionally, theoretical analysis revealed that differences in helical propulsion could induce wobbling in the cell, and that the swimming speed is positively correlated with the wobbling effect. This observation suggests a potential control strategy for helical microswimmers: by generating a lateral force in a two-helix system, it may be possible to alter the swimming direction without the need for altering the source of propulsion, such as an external magnetic field.

However, our findings are limited to a system involving two constant-speed helices in a Newtonian fluid and a simplified 2D theoretical framework. Further research and a more comprehensive theory are required to fully understand the complex interactions among multiple flagella in non-Newtonian fluids.

\clearpage

\begin{figure}[htbp] 
    \centering
    \includegraphics[width=1\linewidth]{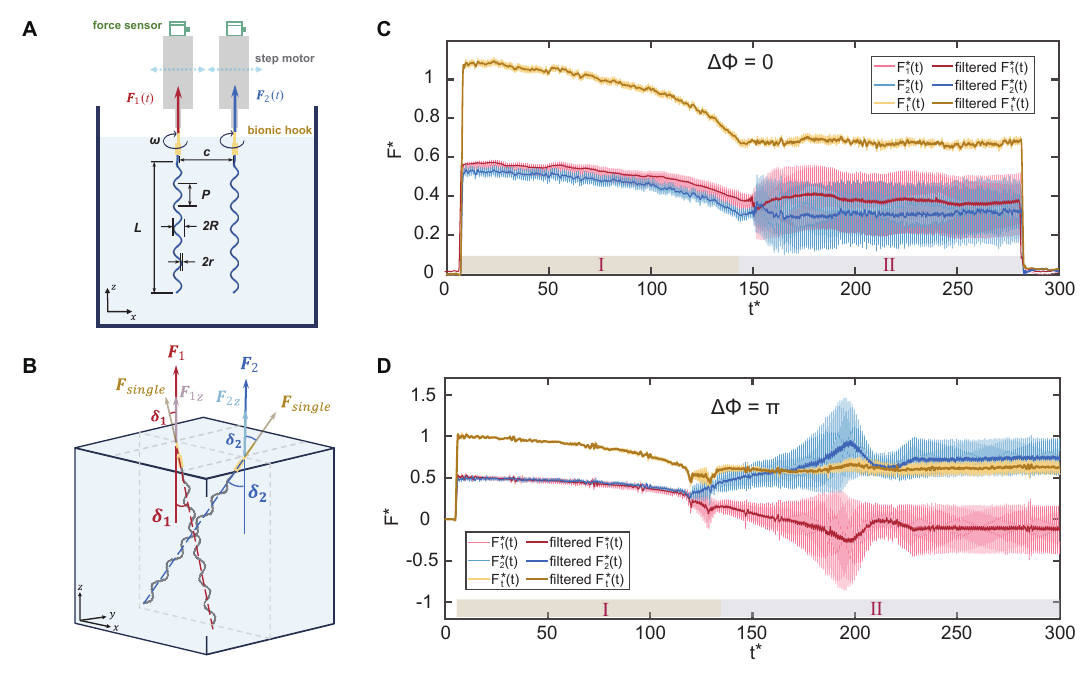}
\end{figure}

\noindent {\bf Fig. 1.} (\textbf{A}) Illustration of macroscopic experiment system from both sides of views, where $L$, $c$, $P$, $\omega$, $R$, $r$ represent helix length, separation distance, helical pitch, rotation speed, helical radius, and filament radius separately (details see SI section 1.1). (\textbf{B}) Illustration of inclination angle ($\delta_{i}$), preconditioned single helix propulsion force ($F_{single}$), propulsion excluding hydrodynamic interaction ($F_{i,z} = F_{single} cos(\delta_i)$) in the bundling process. (\textbf{C, D}) Measured propulsion force trend during bundling process with fixed separation distance ($c$ = $73.6$ mm) and two phase difference ($\Delta \Phi$ = $0$ rad and $\Delta \Phi$ = $\pi$ rad). The figures illustrate the trends in the scaled force $F^*_{1}(t)$, $F^*_{2}(t)$ and $F^*_{t}(t)$ over scaled time ($t$). The thinner lines represent the original data for $F^*_{1}(t)$, $F^*_{2}(t)$ and $F^*_{t}(t)$, while the thicker lines indicate the filtered data, showcasing the effect after filtering the frequency of rotation speed ($\omega$). The helices bundling process manifests in two distinct states: a bundling state (state \uppercase\expandafter{\romannumeral1}) and a steady state (state \uppercase\expandafter{\romannumeral2}), as lower color lump shows.

\clearpage

\begin{figure} 
    \centering
    \includegraphics[width=1\linewidth]{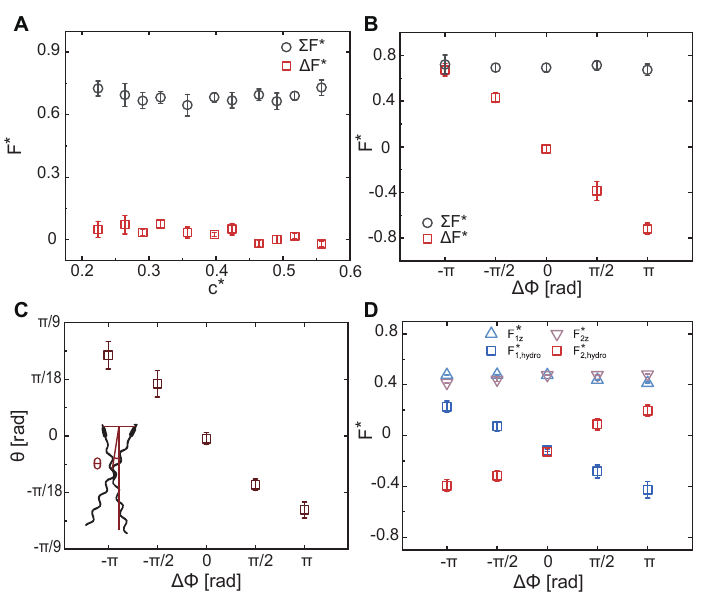}
\end{figure}

\noindent {\bf Fig. 2.} Influence of separation distance ($c^*$) and phase difference ($\Delta \Phi$) to steady bundling system. (\textbf{A}) Scaled force summation ($\Sigma F^*$) and difference ($\Delta F^*$) as a function of separation distance ($c^*$). Each data point represents 15 independent measurements from more than 30s in steady state in the bundling process. (\textbf{B}) Scaled force summation ($\Sigma F^*$) and difference ($\Delta F^*$) as a function of phase difference ($\Delta \Phi$). (\textbf{C}) Leaning angle ($\theta$) as a function of phase difference ($\Delta \Phi$). Leaning angle ($\theta$) is defined as the small angle between the $z$ axis and the line connecting the intersection of the helices and the origin as the inset shows. (\textbf{D}) Comparison between propulsion without considering the hydrodynamics interaction ($F^*_{i, z}$) and hydrodynamic interaction ($F^*_{i, hydro}$) under conditions of different $\Delta \Phi$. Each data point in \textbf{B} to \textbf{D} represents $\geq$ 5 independent measurements from more than 30s in steady state in the bundling process.

\clearpage

\begin{figure} 
    \centering
    \includegraphics[width=1.0\linewidth]{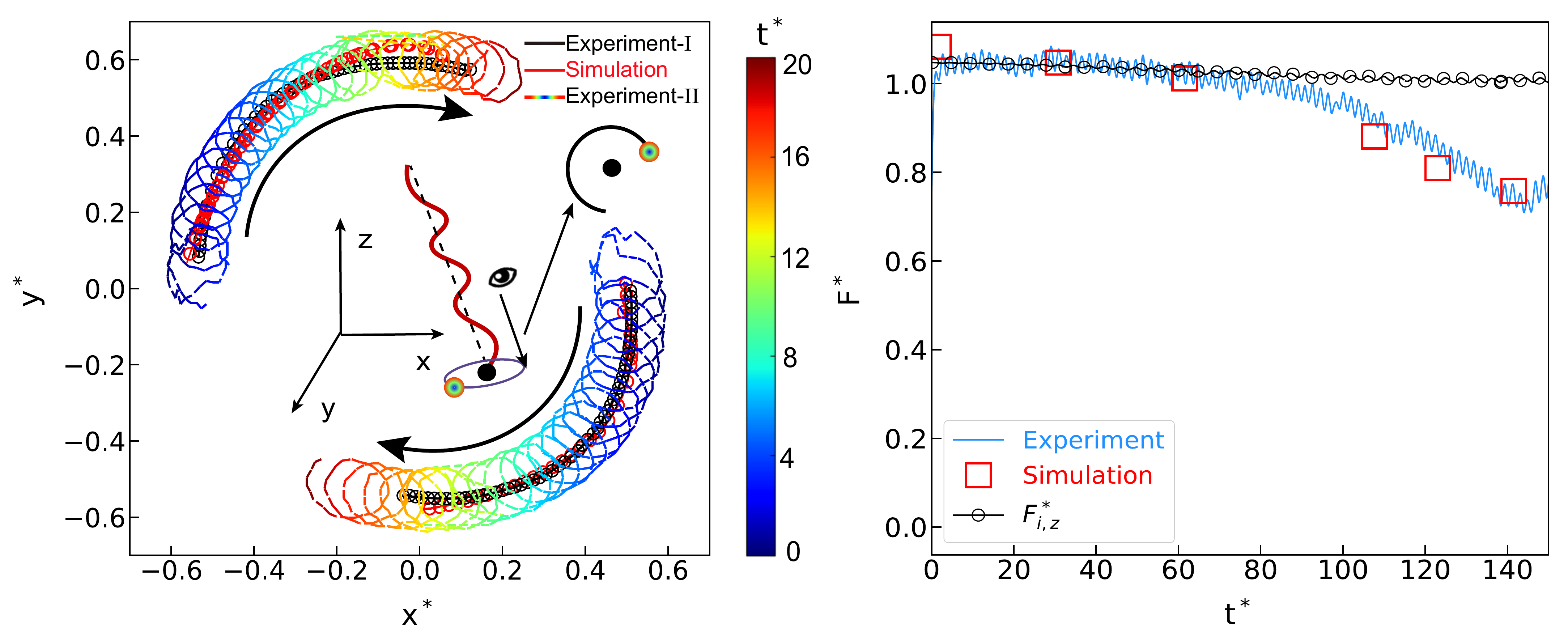}
\end{figure}

\noindent {\bf Fig. 3.} Comparison between simulation results and experimental results. (\textbf{A}) Simulation and experimental flagella tail trajectory comparison with fixed separation distance ($c$ = $39.6$ mm). Red line represent the simulation trajectory (the record point is set on the helical center) ; Black line represent the experimental trajectory (the record point is set on the helical center)  ; The gradient color line represent the experimental trajectory(the record point is set on the flagella tail which is on the same plane as helical center at $t^{*}=0$) ; The position of $x$ and $y$ are scaled as $x^* = x/L$ and $y^* = y/L$ separately. (\textbf{B}) Comparison among experiment, simulation and theoretical results of total propulsion force ($F_t^*$) as a function of time ($t^*$) with fixed separation distance ($c$ = $73.6$ mm) ; Theoretical propulsion force is calculated by $F^*_{i, z} = F^*_{single}cos(\delta_{i})$.  $F^*_{single}$ is the single flagella numerical propulsion (see SI fig. S16).

\clearpage

\begin{figure}[htbp] 
    \centering
    \includegraphics[width=0.8\linewidth]{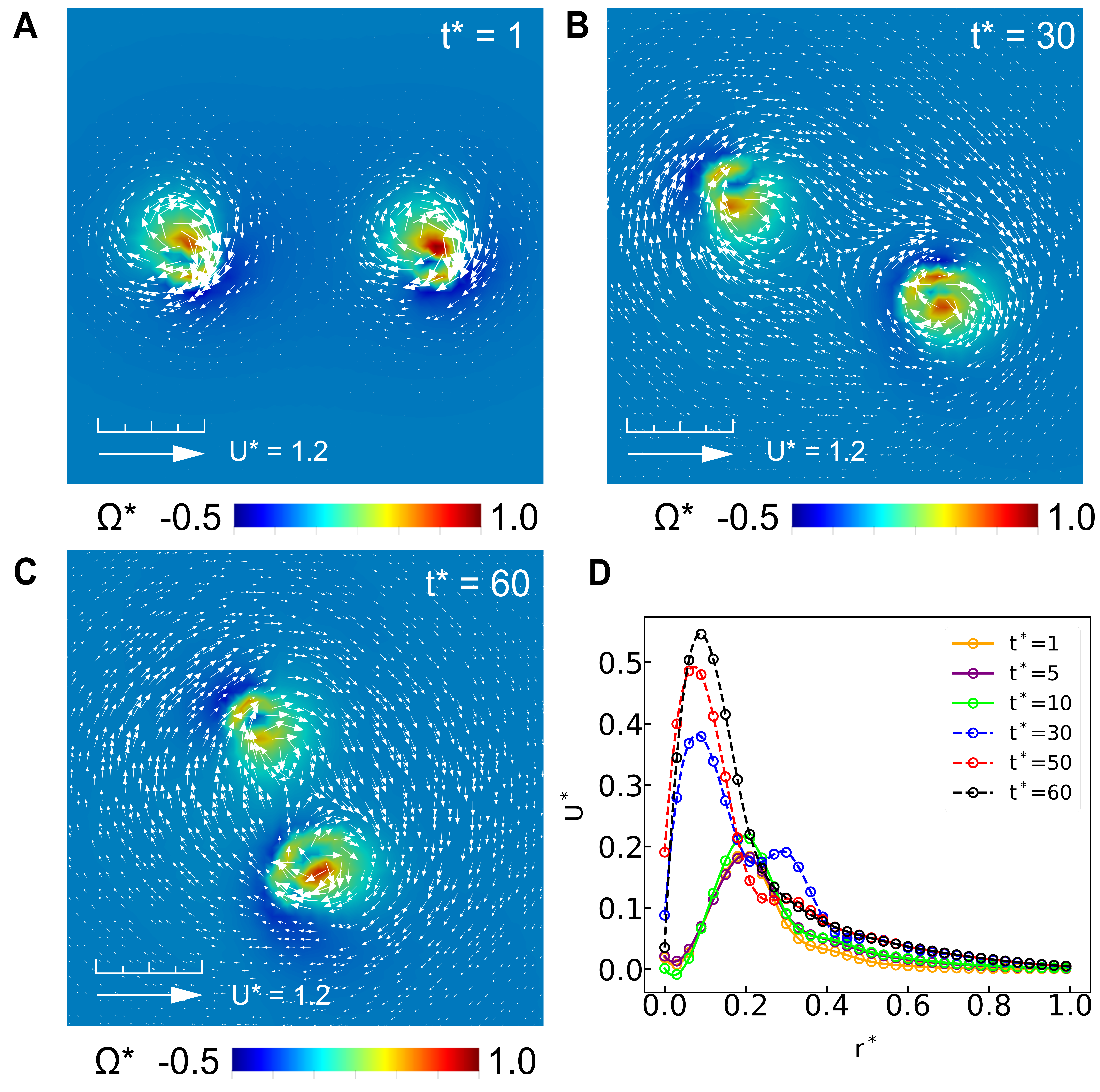}
\end{figure}

\noindent {\bf Fig. 4.} Flow field at instantaneous time in bundling process. (\textbf{A-C}) Vorticity field with instantaneous velocity field at different times. The vorticity in the flow field are scaled as $\Omega^{*} = \Omega/\Omega_{max}$ , where $\Omega_{max}$ is the max vorticity at each instantaneous $t^{*}$. Length of the arrow represents the magnitude of the fluid velocity. Direction of the arrow represents the direction of fluid velocity. The legend is the velocity magnitude represented by the scaleplate. (\textbf{D}) Radial decay of the flow velocity $U^*$ at each instantaneous $t^{*}$, with $r^* = 0$ corresponding to the center of the fluid field, $r^* = 0.2$ corresponding to the flagella rotation axis. Here $U^*$ is normalized by $U^* = U/(\omega R)$ ($U$ is the instantaneous velocity magnitude).

\clearpage

\noindent {\bf Table. 1.} Magnitude of cell speed ($V$) and wobble angle ($2\beta$) with different $\Lambda$
\begin{table}[htbp]
  \centering
  \begin{tabular}{l|ccccc}
    \toprule
    $\Lambda$& $1/35$& $1/5$& $1$& $4$& $68$\\
    \midrule
    $V$ [$\mu$m/s]& $51$& $43$& $40$& $43$& $48$\\
    $2\beta$ [$^\circ$]& $42.16$& $22.9$& $0.6$& $21.0$& $37.1$\\    
    \bottomrule
  \end{tabular}
  \label{Table::simuResult}
\end{table}

\clearpage

\bibliography{scibib}

\bibliographystyle{Science}

\clearpage
\section*{Acknowledgments}
We thank Shuo Guo and Hepeng Zhang for offering many comments for revising the paper. This work is funded by NSFC 12202275, 12372264, STCSM 22YF1419800 and Natural Science Foundation of Shanghai (Grant No. 23ZR1430800).

\section*{Supplementary materials}
Materials and Methods\\
Supplementary Text\\
Figs. S1 to S19\\
Tables S1 to S2\\
References \textit{(46-53)}


\end{document}


\maketitle 
\section{Materials and Methods}
\subsection{Experimental Setup}
\begin{figure}[htbp] 
    \centering
    \includegraphics[width=1\linewidth]{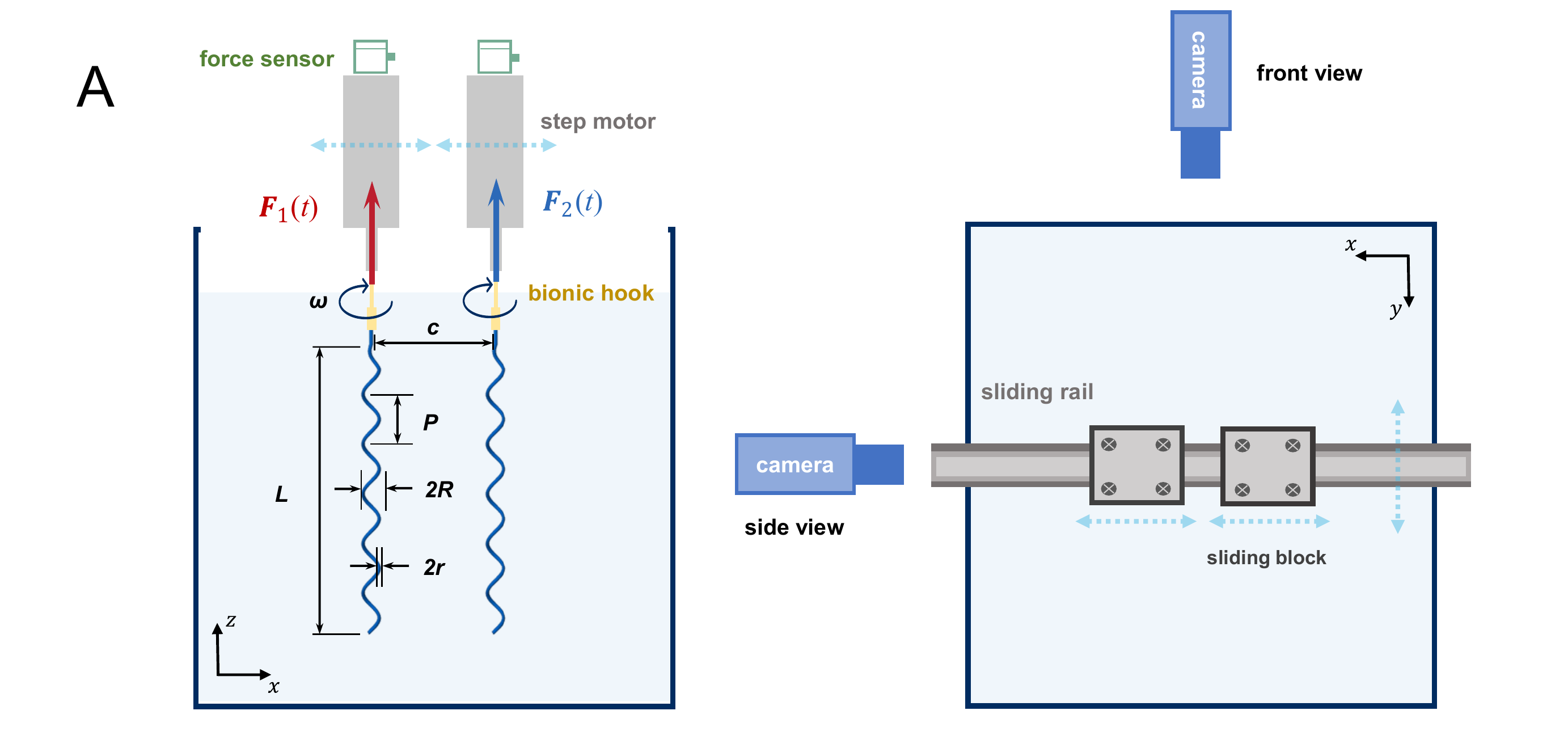}
    \caption{Illustration of macroscopic experiment setup}
    \label{SIfig::setup}
\end{figure}

As shown in the fig.\ref{SIfig::setup}, which is a supplement to Fig. 1, the experimental system employs a specially designed scale model with two cameras (DAHENG IMAGING MER2-302-56U3M) collecting the data of the bundling process from two directions which are marked as $x$ and $y$ respectively. Two identical step motors (UMot 40*20, reduction ratio: 1:139) are used in the experiment. The helices are made of photosensitive resin (details in section 1.2) with geometries given in Table. ~\ref{SITable::helix_parameter}. The motor and the helix are connected by a bionic hook which bends during the bundling process. Two strain-gauge type force sensors (RIGHT T313 Tension \& Compression Force Sensors, maximum tension, $\pm$ 2.5 N; manufacturer’s stated accuracy, $\pm$ 0.08\% of full scale) are attached above the motors to measure the propulsion force over the experiment. The force data are collected by a data acquisition device (National Instruments USB-6211) driven by LabVIEW in high frequency (2000 Hz). Both cameras are synchronized using Galaxy Windows SDK provided by DAHENG IMAGING. To minimize the boundary effects, all helices are centrally located in the tank (300mm $\times$ 300 mm $\times$ 300mm) to minimize the wall effect. Additionally, macroscopic helical models with bionic hooks are designed to mimic the actual geometry of microorganism flagella. The helix is fabricated by 3D printing from photosensitive resin. The geometric parameters of the helix is given in Table. ~\ref{SITable::helix_parameter}.  This experiment applied bionic hook structure ($E_{hook} \approx $ 600 MPa, estimated based on information provided by manufacturer), enabling the rigid filament to change its axial direction while rotating. 

\begin{table}[htbp]
 \caption{Helix model parameters}
  \centering
  \begin{tabular}{l|c}
    \toprule
    Parameter & Value \\
    \midrule
    Length ($L$) & 150 mm \\
    Helical pitch ($P$) & 28.3 mm \\
    Helical radius ($R$) & 4.5 mm \\
    Filament radius ($r$) & 1.5 mm \\
    Young's Modulus ($E$) & 2.4 GPa \\
    \bottomrule
  \end{tabular}
  \label{SITable::helix_parameter}
\end{table}

\begin{figure}[htbp] 
    \centering
    \includegraphics[width=0.6\linewidth]{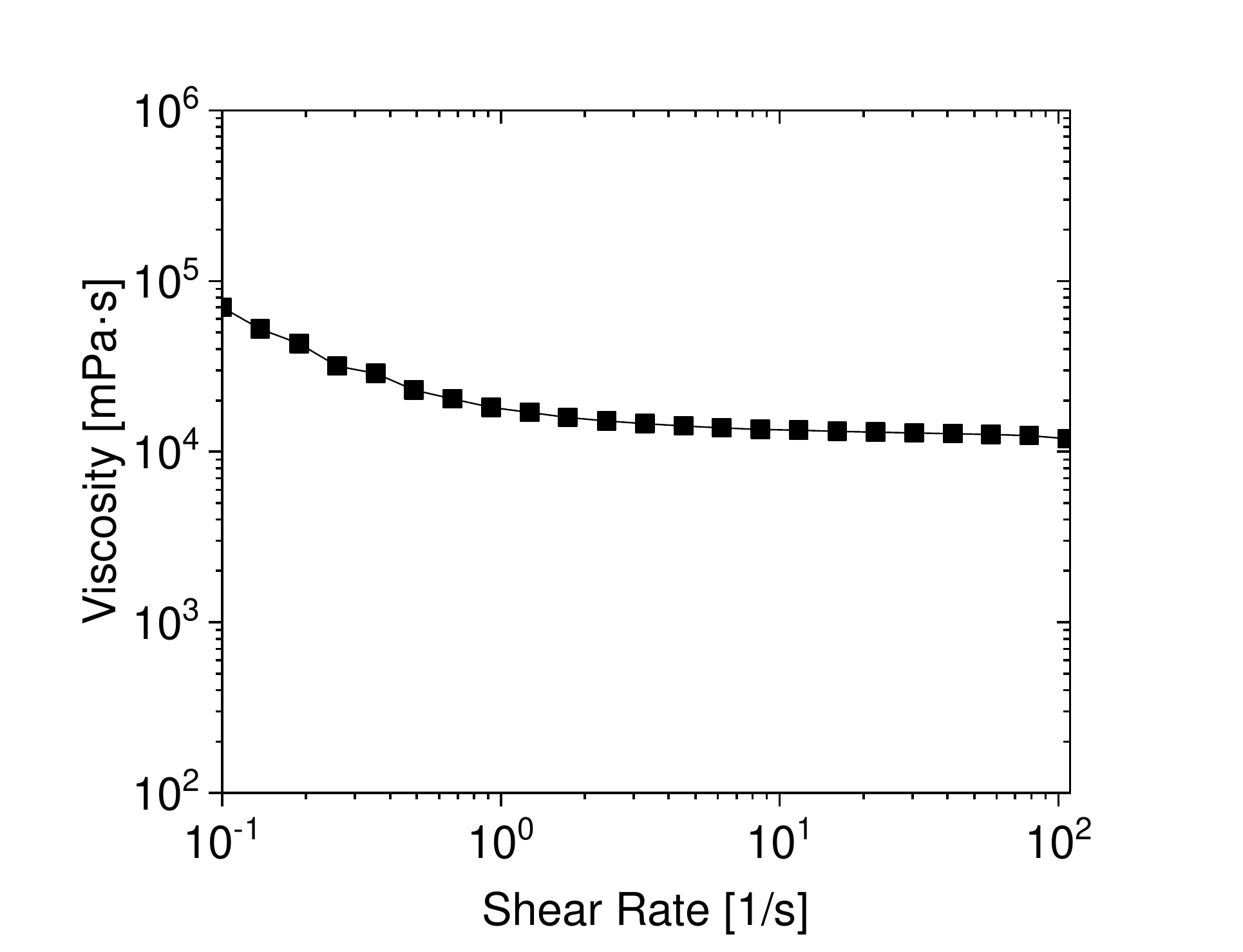}
    \caption{Viscosity of silicon oil used in the experiment as a function of shear rate.}
    \label{SIfig::rheology}
\end{figure}

The fluid used in the experiment is silicon oil which is known to be Newtonian. The viscosity of the silicon oil is about 12.5 N$\cdot$/m$^{3}$ which is measured by a cone-plane shearing rheometer (Anton Paar MCR 302e), which is shown as in fig. ~\ref{SIfig::rheology}. The density is $\rho =$ 0.96 $\times$ 10$^3$ $\mathrm{kg/m^3}$. In this work, the motors always work at a constant angular speed, $\omega =$ 5.08 $\mathrm{rad/s}$. Thus, the \emph{Reynolds} number (\emph{Re}), defined as $Re = 2\rho \omega R r/\eta$, is on the order of $10^{-2}$. The \emph{Sperm} number (\emph{Sp}), which measures the ratio between the viscous force and elastic force during the bacterial flagella bundling process, is defined as $Sp = (\frac{4\eta \omega L^4}{\pi E r^4} )^\frac{1}{4}$ \cite{KIM2003, PhysRevFluidsQu, Machin1963}. And $Sp$ number in this work is $3.21$. This is consistent with the former studies \cite{KIM2003, PhysRevFluidsQu, Man2017}. Therefore, we believe our macroscopic model is dynamically similar to microscopic systems like bacterial flagella \cite{turner_real-time_2000,Darnton_Turner} or helical microswimmers \cite{zhang_characterizing_2009,ahmed_artificial_2016,yan_multifunctional_2017,choi_recent_2021,deng_acoustically_2023}.

\subsection{Fabrication Procedure}
\begin{figure}[htbp] 
    \centering
    \includegraphics[width=1\linewidth]{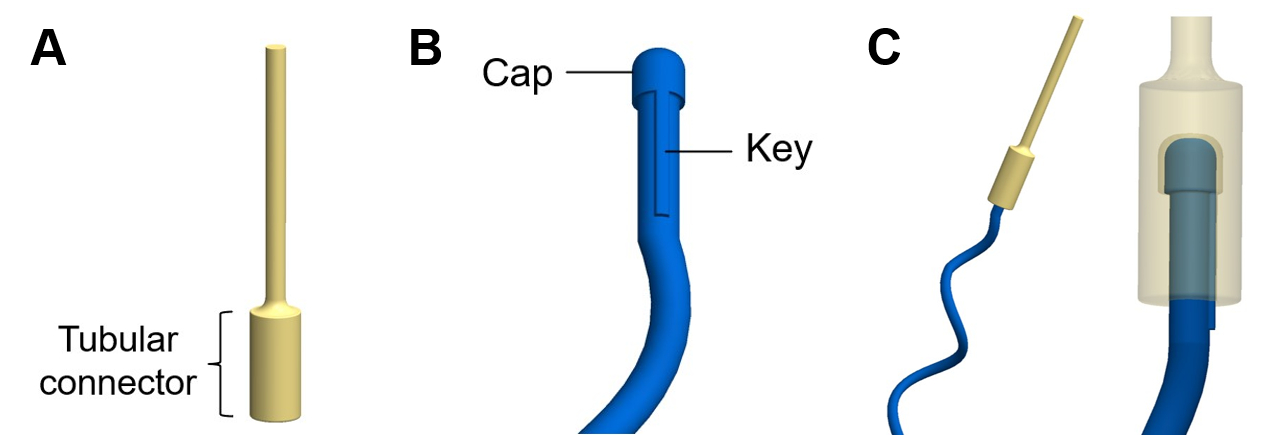}
    \caption{Modified version of bionic helix model with "hook" structure. (\textbf{A}) "Hook" structure. (\textbf{B}) modified filament (top part). (\textbf{C}) assembled flagella model.}
    \label{SIfig::hook}
\end{figure}

Compared to the rigid helix, the bionic hook has more elasticity and flexibility. This soft segment is manufactured by 3D printing from soft rubber with more elasticity (Wemax). This soft segment enables the rigid filament to change its axial direction flexibly. Inspired by this, we designed a similar "hook" made of soft rubbers and connected it onto the modified filament model (The cap and key on the top of filament enable their connection without slipping). fig. ~\ref{SIfig::hook} shows the "hook" structure and the modified version of flagella model. 

\begin{figure}[htbp] 
    \centering
    \includegraphics[width=1\linewidth]{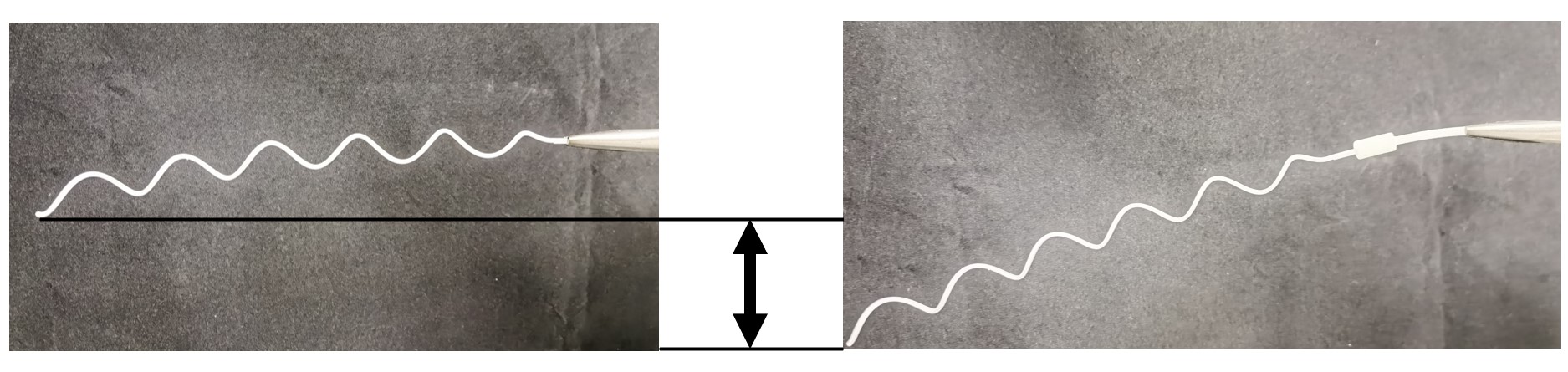}
    \caption{ Increased deflection under same loading condition (left: original version; right: modified version)}
    \label{SIfig::deformation}
\end{figure}

We tested the flexibility of the models before and after the modification, with results shown in fig. ~\ref{SIfig::deformation}. It turned out that the deformation of the new model is much more apparent than the previous one when they are under the same loading condition. Finally, we did some experiments with our new prototype and successfully obtained the bundling effect, which verifies that the hydrodynamic property of our bionic helix model is comparable with these real flagella. Hence, this model is eligible for the macroscopic study of flagella interaction and propulsion generation. 

\clearpage
\section{Supplementary Text}
\subsection{Supplementary Text: Experimental results}
\subsubsection{Resistive Force Theory (RFT)}
Resistive force theory calculates the overall force and torque exerted during the motion of a flagellum by summing the local forces acting on each small segment of the flagellum based on previous works \cite{gray1955propulsion,lighthill1976flagellar,rodenborn_propulsion_2013}.  Here, the pitch angle $\theta$ is defined as $arctan(\frac{P}{2\pi R})$. 
\begin{equation}
    F=(\omega R)\left(C_n-C_t\right) \sin \theta \cos \theta \frac{L}{\cos \theta} \\
\end{equation}
With Gray and Hancock’s drag coefficients are:
\begin{equation}
C_t=\frac{2 \pi \mu}{\ln \frac{2 P}{r}-1 / 2} \quad \text { and } \quad C_n=\frac{4 \pi \mu}{\ln \frac{2 P}{r}+1 / 2}
\end{equation}
Based on our geometry parameters and dimensionless process, our propulsion force can be computed as: $F_{RFT}^* = 0.66$. Comparing to the measured value in the experiment, $F^*_{single} \approx 0.55$, the experimental results can be considered as reasonable.

\subsubsection{Data processing for experimental results}
\begin{figure}[htbp] 
    \centering
    \includegraphics[width=1\linewidth]{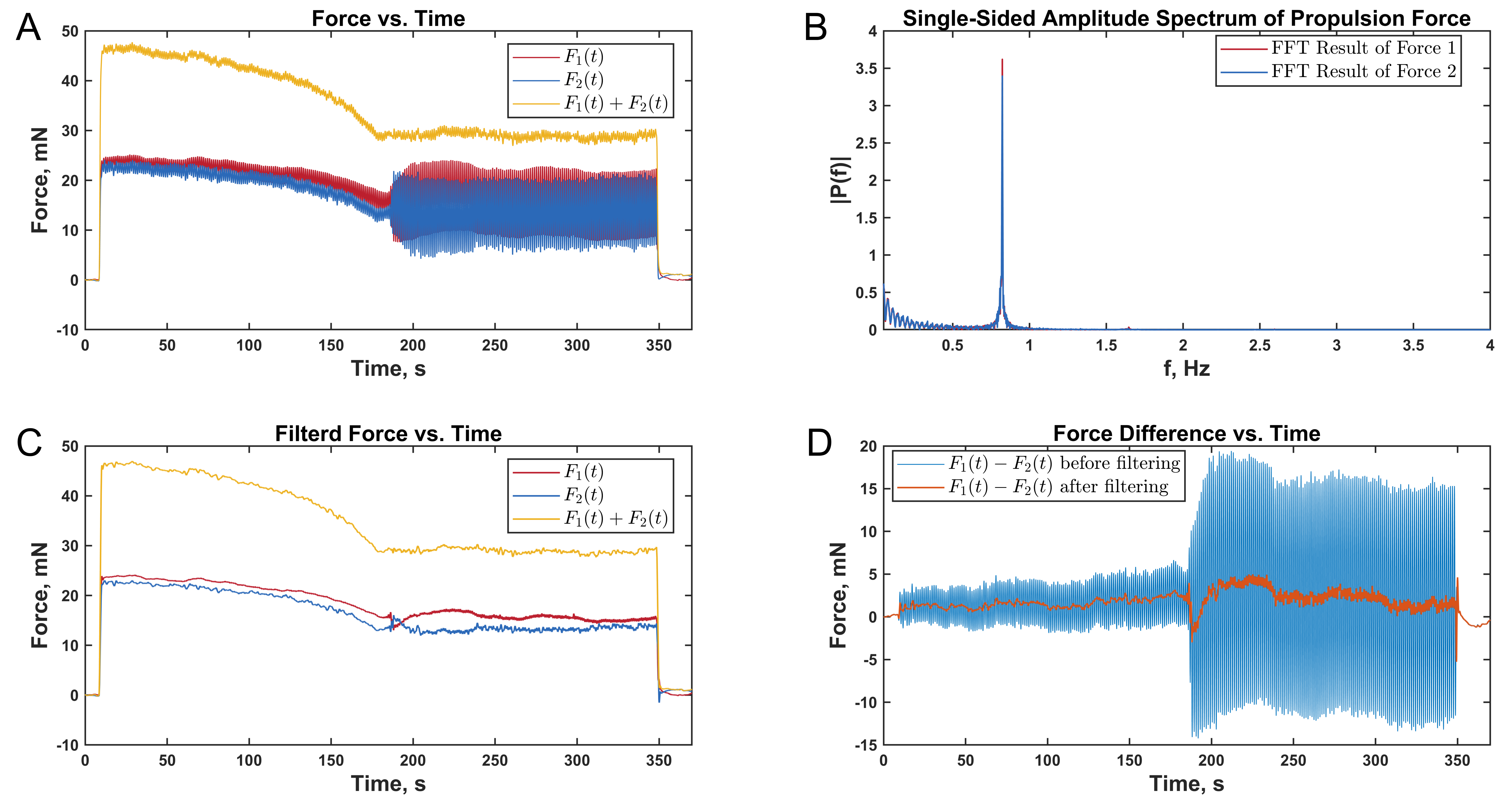}
    \caption{Data analysis processed by MATLAB. (\textbf{A}) Smoothed result of propulsion force ($F_1(t)$, $F_2(t)$, and $F_1(t) + F_2(t)$) as a function of time ($t$). (\textbf{B}) FFT result of propulsion force vs. time in \textbf{A}. Dominant frequencies of $F_1(t)$ and $F_2(t)$ are identical as $\omega_{dominant} = 0.81$. (\textbf{C}) Propulsion force ($F_1(t)$, $F_2(t)$, and $F_1(t) + F_2(t)$) as a function of time ($t$) after bandwidth filtering the $\omega_{dominant}$ in \textbf{B}. (\textbf{D}) Comparison between force difference $F_1(t) - F_2(t)$ as a function of time ($t$) before and after filtering.}
    \label{SIfig::analysisProcess}
\end{figure}

The data processing method in this work is illustrated as fig. ~\ref{SIfig::analysisProcess}. The raw propulsion force is firstly smoothed by method of "Gaussian" smooth algorithm in MATLAB with a 2000 width window. Then, fast Fourier transform is applied to the smoothed data and get the dominant frequency, which is tested to be identical to the rotation frequency of motors. The influence of helices rotation with a frequency of 0.81 Hz is filtered out then, getting results showing in the filtered force. Also, the same method is applied to force difference between helices, getting a result of $F_1(t)-F_2(t)$ before and after filtering as fig. ~\ref{SIfig::analysisProcess} shows.

\subsubsection{Experiment results: Propulsion Force vs. rotation speed}
\begin{figure}[htbp] 
    \centering
    \includegraphics[width=0.6\linewidth]{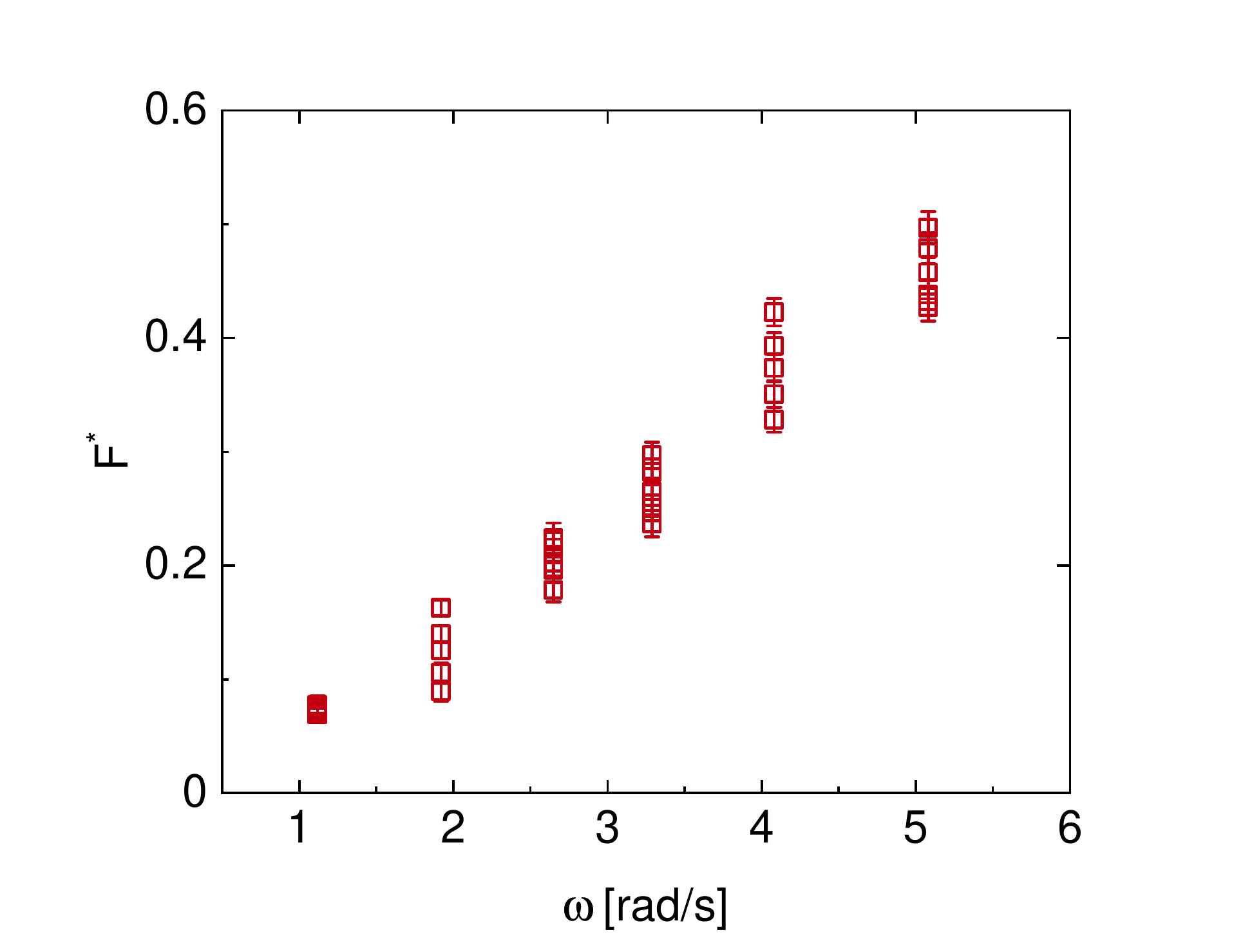}
    \caption{Single helix propulsion force ($F_{single}^*$) as a function of rotation speed ($\omega$).}
    \label{SIfig::forcevsrotationspeed}
\end{figure}

This section shows some primary results showing the relationship between propulsion force vs. rotation speed. As fig. ~\ref{SIfig::forcevsrotationspeed} shows, the propulsion of single helix ($F^*_{single}$) increases linearly as the increment of rotation speed ($\omega$). In the RFT prediction as previous section introduces, $F=(\omega R)\left(C_n-C_t\right) \sin \theta \cos \theta \frac{L}{\cos \theta} $, which shows that the propulsion force is linear correlated to rotation frequency.

\clearpage
\subsubsection{Experiment results: difference phase inclination example}
\begin{figure}[htbp] 
    \centering
    \includegraphics[width=0.8\linewidth]{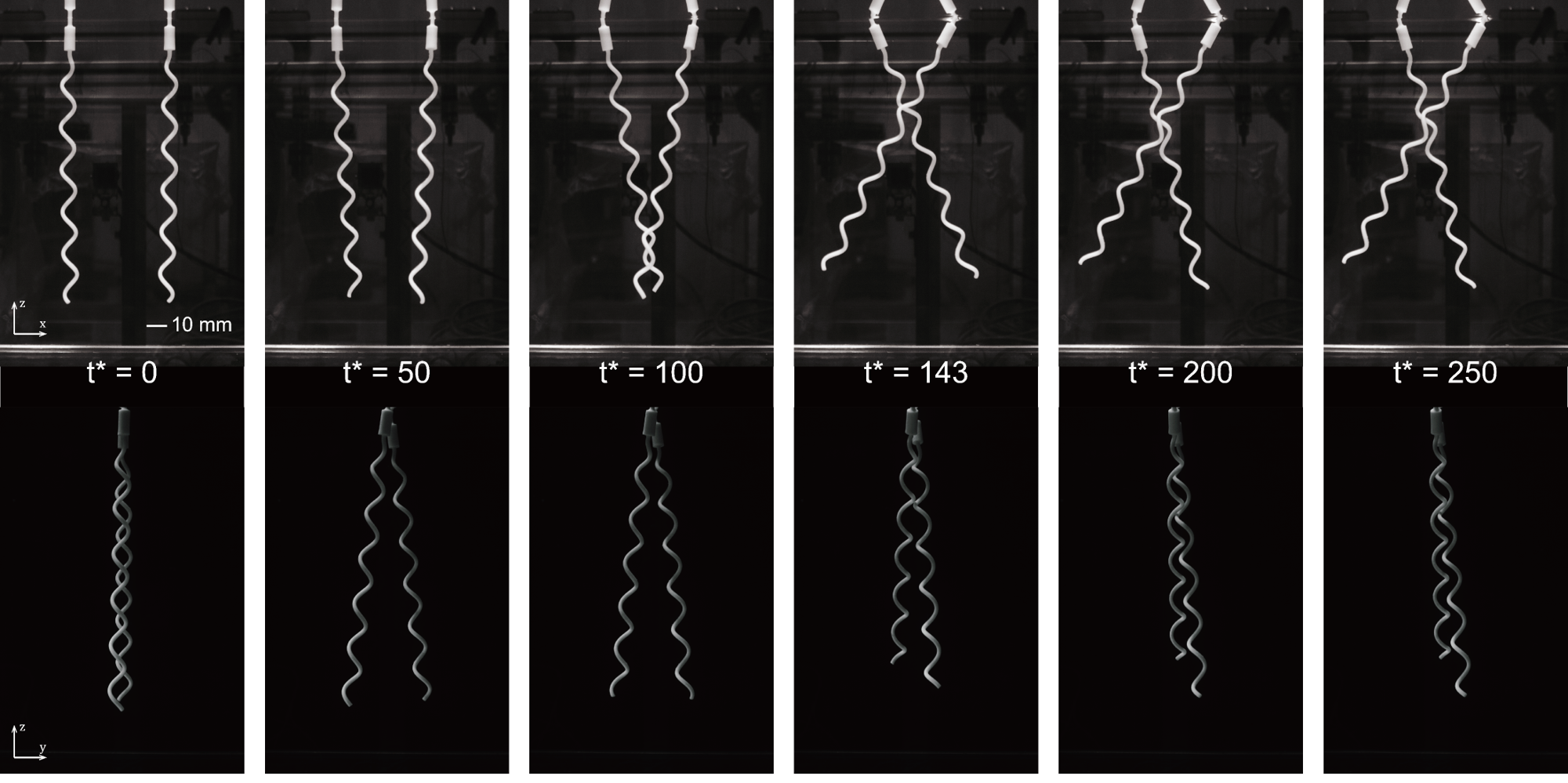}
    \caption{Bundle process when $\Delta \Phi = -\pi$}
    \label{SIfig::bundleprocess_minuspi}
\end{figure}

\begin{figure}[htbp] 
    \centering
    \includegraphics[width=0.8\linewidth]{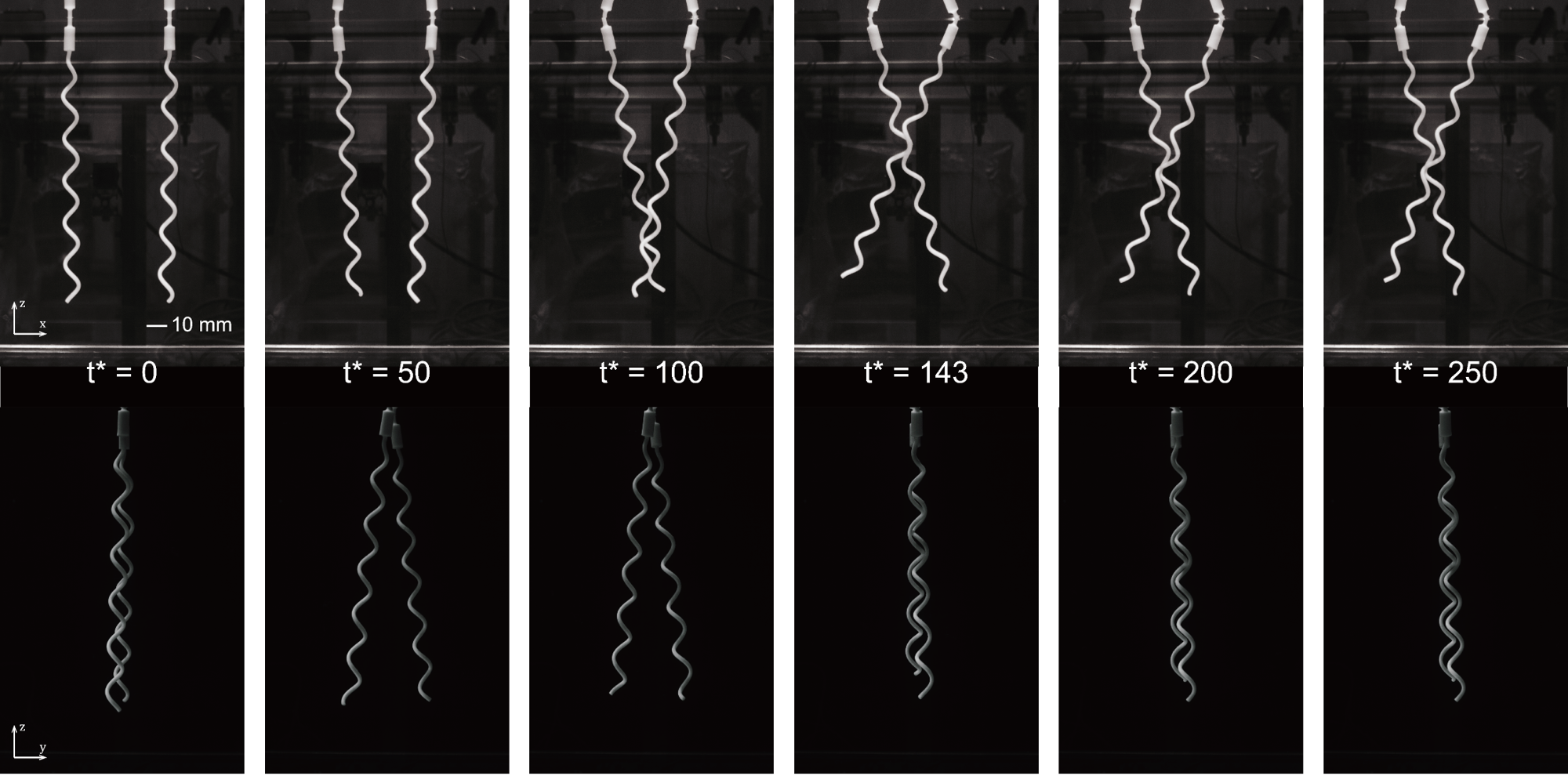}
    \caption{Bundle process when $\Delta \Phi = -\pi/2$}
    \label{SIfig::bundleprocess_minuspi/2}
\end{figure}

\begin{figure}[htbp] 
    \centering
    \includegraphics[width=0.8\linewidth]{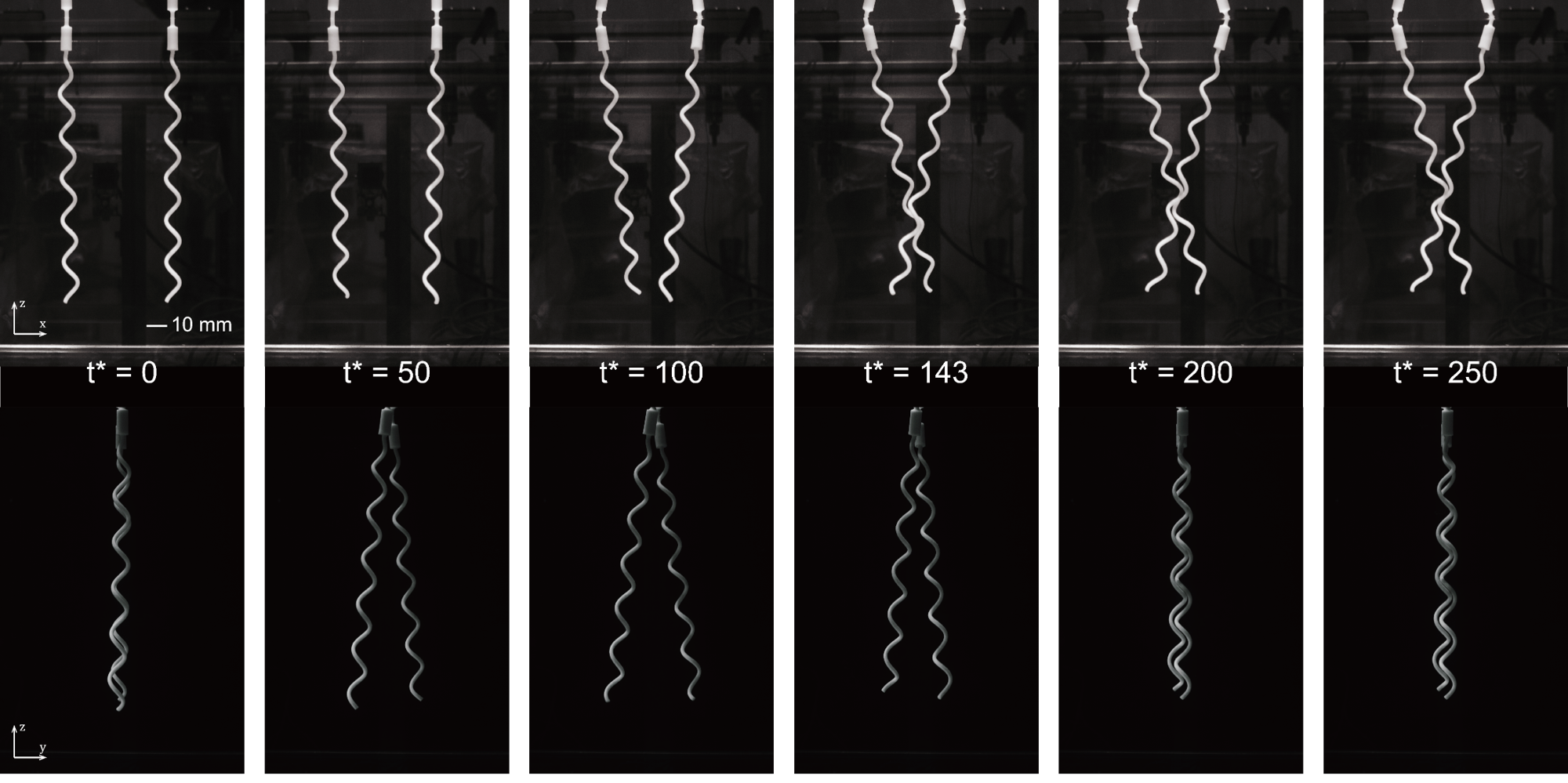}
    \caption{Bundle process when $\Delta \Phi = 0$}
    \label{SIfig::bundleprocess_0}
\end{figure}

\begin{figure}[htbp] 
    \centering
    \includegraphics[width=0.8\linewidth]{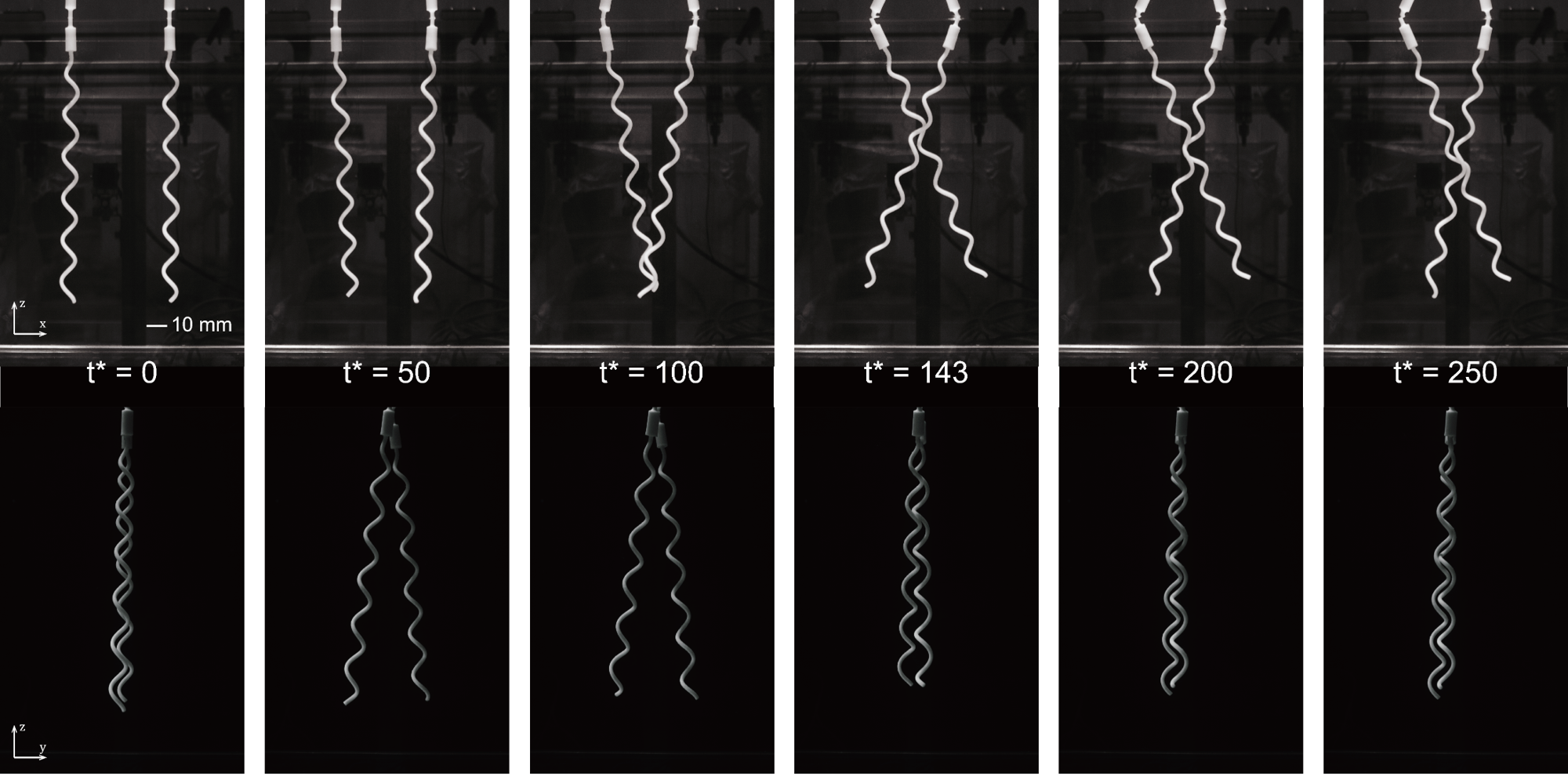}
    \caption{Bundle process when $\Delta \Phi = \pi/2$}
    \label{SIfig::bundleprocess_pi/2}
\end{figure}

\begin{figure}[htbp] 
    \centering
    \includegraphics[width=0.8\linewidth]{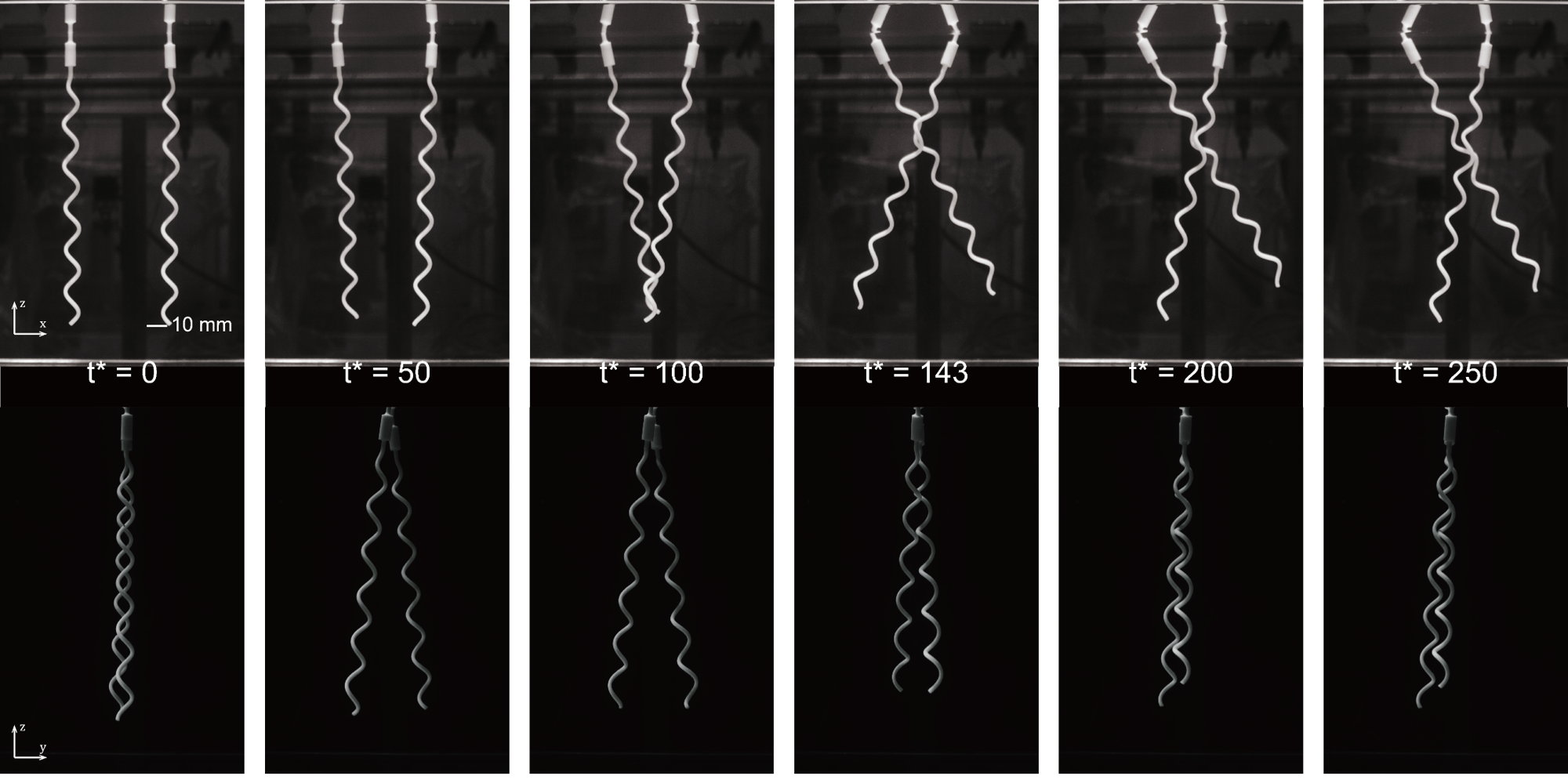}
    \caption{Bundle process when $\Delta \Phi = \pi$}
    \label{SIfig::bundleprocess_pi}
\end{figure}

This section includes several typical bundling process with identical separation distance ($c$) of 73.6 mm but with different initial phase difference ($\Delta \Phi)$. 
Case of $\Delta \Phi = 0$ showing in fig. \ref{SIfig::bundleprocess_0} is synchronized to the force trend presented in the main text.  Also, fig. ~\ref{SIfig::bundleprocess_minuspi}, fig. ~\ref{SIfig::bundleprocess_minuspi/2}, fig. ~\ref{SIfig::bundleprocess_pi/2} and fig. ~\ref{SIfig::bundleprocess_pi} illustrate the bundling process with different phase difference as $\Delta \Phi = -\pi$, $\Delta \Phi = -\pi/2$, $\Delta \Phi = \pi/2
$, and $\Delta \Phi = \pi$ separately. Combining the images sequence showing the bundle process and the computation of leaning angle presented in the main text, the change of $\Delta \Phi$ leads to the symmetry loss of the steady bundle system. And thus, the helices bundle leans towards the helix with a behindhand phase to maintain in-phase state. Also, the influence of phase difference $\Delta \Phi$ is symmetric at the case of $\Delta \Phi = 0$. This phenomenon is coherent with the results of both leaning angle result and propulsion results.


\subsubsection{Supplement to the influence of separation}

\begin{figure}[htbp] 
    \centering
    \includegraphics[width=1\linewidth]{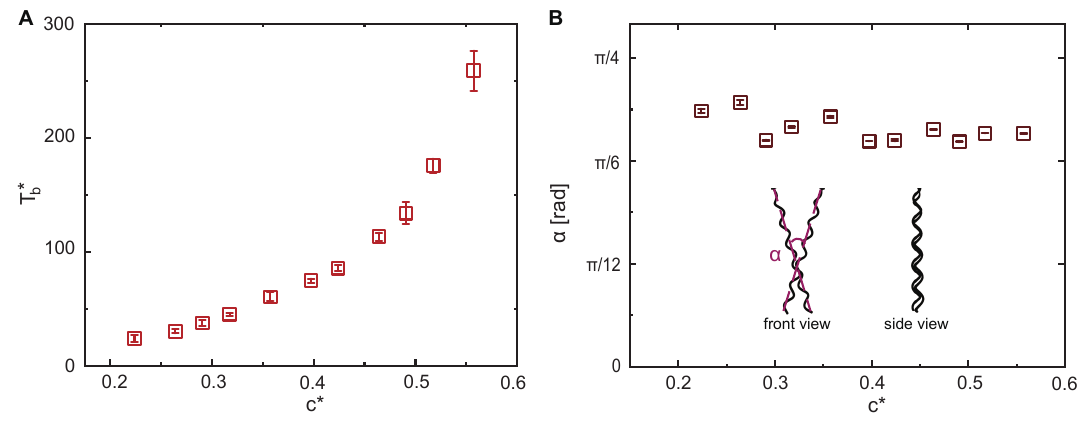}
    \caption{Supplement to the influence of separation distance ($c^*$). (\textbf{A}) Measured bundling time ($T_{b}^*$) as a function of $c^*$. (\textbf{B}) Included angle ($\alpha$) as a function of $c^*$. Inset in \textbf{B} shows the definition of $\alpha$, which is defined as  included angle between the axes of two helices in the x-z plane, considering very slight inclination in the y-z plane.}
    \label{SIfig::separtion_influence}
\end{figure}

The measured bundling time, $T_{b}^{*}$ = $\omega T_b /(2\pi)$ where $T_{b}$ is defined as the time when the system reaches the steady state, is ploted over $c^{*}$ in fig. ~\ref{SIfig::separtion_influence}\textbf{A}. Clearly, $T_{b}^{*}$ increases with respect to $c^*$ as expected considering the equilibrium between hydrodynamic and elastic forces in the bundling process \cite{kim_hydrodynamic_2004,Man2017, reichert_synchronization_2005,lim_bacteria-inspired_2023}. This also partly explains the significant variation in the experimentally measured bundling time in the former studies \cite{berg_chemotaxis_1972,patteson_running_2015,qu_changes_2018}.  Similar to propulsion force, the included angle ($\alpha$) between two individual helices at stable state is unaffected by the separation distance between helices. $\alpha$ is obtained by measuring the included angle between the axes of two helices as fig S. \ref{SIfig::separtion_influence}\textbf{B} illustrates. In the steady state, the helices remain stable rotation in the the x-z plane defined in fig S. \ref{SIfig::systemSetup}, which can been seen in the inset of side view in fig S. \ref{SIfig::separtion_influence}\textbf{B} . Previous works mentioned the equilibrium between hydrodynamic and elastic forces in the bundling process of helices \cite{Man2017, lim_bacteria-inspired_2023}. With approximately equalized $\alpha$, the angle between each helix and the x-z plane shows no distinct difference. Thus, the elastic force caused by bending of the hook \cite{Man2017} is considered at a similar magnitude in this experiment. Consequently, the hydrodynamic interaction between helices is not influenced by the separation if $c^* < 1$. Based on slender-body theorem, it is known that the influence of helix exerting to surrounding fluid decreases with distance non-linearly \cite{lighthill1976flagellar}. In this case, the enlarging separation distance leads to smaller hydrodynamic force on the helices. Since the elastic force in the stable state remains constant,  the time to reach equilibrium between hydrodynamic force and elastic force increases non-linearly. 

\subsubsection{Supplement to the influence of phase difference}
As a addition to Fig. 2 in the main text, we showed more detailed measured force as the following Table. ~\ref{SITable::F1&F2} shows:
\begin{table}[htbp]
 \caption{Supplement to measured propulsion force, $F^*_1$ \& $F^*_2$ and inclination angle of individual helix, $\delta_1$ \& $\delta_2$.}
  \centering
  \begin{tabular}{l|ccccc}
    \toprule
    $\Delta \Phi$ [rad]& $-\pi$ & $-\pi/2$ & $0$ & $\pi/2$  & $\pi$\\
    \midrule
    $F^*_1$ & $0.70 \pm 0.04$ & $0.56 \pm 0.04$ & $0.36 \pm 0.01$ & $0.16 \pm 0.05$& $-0.01 \pm 0.06$\\
    $F^*_2$ & $0.02 \pm 0.05$ & $0.12 \pm 0.04$ & $0.35 \pm 0.03$ & $0.57 \pm 0.05$ & $0.68 \pm 0.04$\\    
    $\delta_1$ [$^\circ$]& $16.1 \pm 0.7$& $15.0 \pm 1.5$& $16.6 \pm 0.2$& $27.9 \pm 1.2$& $33.3 \pm 5.9$\\
    $\delta_2$ [$^\circ$]& $-33.1 \pm 0.3$& $-27.2 \pm 3.8$& $-16.2 \pm 0.1$& $-16.2 \pm 1.0$& $-13.6 \pm 4.5$\\
    \bottomrule
  \end{tabular}
  \label{SITable::F1&F2}
\end{table}

\begin{figure}[htbp] 
    \centering
    \includegraphics[width=0.8\linewidth]{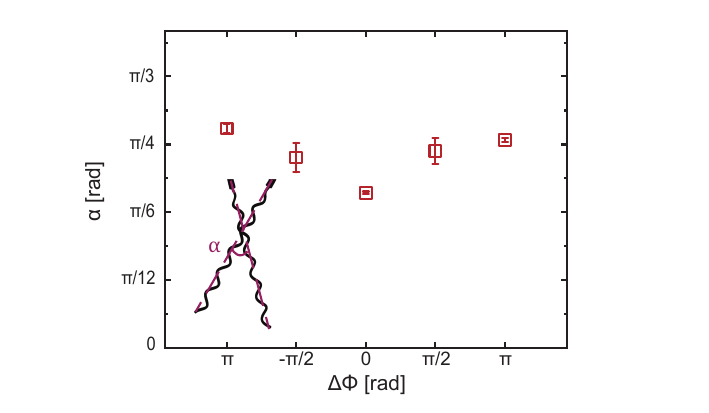}
    \caption{Supplement to the influence of phase difference ($\Delta \Phi$). Included angle ($\alpha$) as a function of $\Delta \Phi$.}
    \label{SIfig::phase_alpha}
\end{figure}

Besides, we also studied the influence of $\Delta \Phi$ to included angle ($\alpha$) between helices when in the steady state in the x-z plane. Here from the fig. \ref{SIfig::phase_alpha}, we can see that when there is no initial phase difference between two helices, the angle between two helices in the stable state is the minimum at about $ \pi/5 $. When in the case of $\Delta \Phi = -\pi$ and $\Delta \Phi = \pi$ initial phase difference, the angle between two helices in the stable state reached maximum at about $ 4\pi/15 $. The slight asymmetry of included angle can be explained by the inconsistency in the manufacturing (3D-printing) of bionic hooks and the helices. 

\newpage

\subsection{Supplementary Text: Numerical method}
The overset numerical solver developed here comprises two subsolvers: (i) the flow solver to solve the N-S equation; and (ii) the lagrangian solid solver to solve the solid body motion equation. A two-step solution strategy is used. First, the three-dimensional(3-D) flow equations around the flagellum body are solved using the pimple algorithm \cite{fluids9020051} to obtain velocity and pressure fields. Velocity and pressure around the body are used to calculate the hydrodynamic force on the surface of the solid body. Second, the forces are passed on to the solid solver, then the solid motion equation is solved to update the position and velocity of the solid part. These steps allow for two-way solid-fluid coupling at every time step. The solver allows for arbitrary number of solids with arbitrary geometric features in six degree of freedom (6 DOF) motion.

\subsubsection{Fluid solver: governing equations and solution methodology}
The fluid is assumed to be incompressible and Newtonian, and the immersed solid is rigid and non-porous. These assumptions are consistent with the experimental conditions. Thus, the governing equations for the fluid flow are:
\begin{equation}
\boldsymbol{\nabla} \cdot \boldsymbol{u}=0
\end{equation}

\begin{equation}
\frac{\partial \boldsymbol{u}}{\partial t}+\boldsymbol{u} \cdot \boldsymbol{\nabla} \boldsymbol{u}=-\frac{1}{\rho_f} \nabla p+\nu \nabla^2 \boldsymbol{u}.
\end{equation}
Here, $\nu$ is the kinematic viscosity of the fluid, $p$ is the pressure, $\rho_f$ is the fluid density.  A no-slip(wall), no-penetration boundary condition is enforced on the solid-fluid interface in viscous flow. For the solid part, we have
\begin{equation}
\boldsymbol{u}_{\Gamma}= \boldsymbol{0}.
\end{equation}
\begin{equation}
\nabla p \cdot \boldsymbol{n}_{\Gamma}=0.
\end{equation}
Here, $\boldsymbol{u}_{\Gamma}$ is the fluid velocity on the solid surface. $\Gamma$ indicates the solid immersed surface and $\boldsymbol{n}_{\Gamma}$ indicates the normal to the solid. 

\subsubsection{Solid body 6-DOF algorithm}
The governing equations for the solid are Newton's second law built on the global coordinates system which will remain stationary during the numerical. While any number of solids can be immersed in the framework, computational effort rises with increasing number of solids. The solid is free to perform full 6DOF motion. Force and moment balance equation are as follows:
\begin{equation}
m \frac{\boldsymbol{d} \boldsymbol{v}}{\boldsymbol{d} t}= \boldsymbol{F}_O
\end{equation}
\begin{equation}
\mathcal{I}_s \frac{\boldsymbol{d} \boldsymbol{\Omega}}{\boldsymbol{d} t}= \boldsymbol{M}_O
\end{equation}
Where the force $\boldsymbol{F}_O$ and the moment $\boldsymbol{M}_O$ is applied by the fluid on the solid-fluid interface(i.e the solid surface). $\boldsymbol{v}$ is the translation velocity. Usually Equation (6) is build in inertial principal axis coordinate system, Equation (5) is build in global frame. $\boldsymbol{\Omega}$ is the angular velocity around the inertial principal axis. $\mathcal{I}_s$ is inertia of the solid body. $O$ is the center of mass. To be more detailed, $\boldsymbol{F}_O$ and $\boldsymbol{M}_O$ are given as follows:

\begin{equation}
\boldsymbol{F}_O=\iint_S(p \boldsymbol{n}+\boldsymbol{\tau}) d S
\end{equation}
\begin{equation}
\boldsymbol{M}_O=\iint_S\left(\boldsymbol{r}_s \times(p \boldsymbol{n}+\boldsymbol{\tau})\right) d S
\end{equation}

In the above formula, $\boldsymbol{r}_s$ is the relative location of that surface point to the mass centre of the solid. $\boldsymbol{f_O} = (p \boldsymbol{n}+\boldsymbol{\tau})$ is the force exerted by the fluid on a unit solid surface area. This unit force is considered to be the weight of the pressure and viscous force. In numerical numericals, force exerted on the solid surface is discretized into several elements, so we give discrete expressions of pressure and viscous force as follows:

\begin{equation}
\iint_S p \boldsymbol{n} d S=\sum_i s_{i} \boldsymbol{n}_{i} p_i
\end{equation}
\begin{equation}
\iint_S \boldsymbol{\tau} d S=\sum_i s_{i} \boldsymbol{n}_{i} \cdot\left(\mu \boldsymbol{\tau}_{\mathrm{dev}}\right)
\end{equation}
$s_{i}$ is the surface element area, $\boldsymbol{n}_{i}$ is the surface element normal vector, $\boldsymbol{\tau}_{\mathrm{dev}}$ is the deviatoric stress tensor exerted on surface element, $\mu$ is the fluid viscosity. $p_i$ is the pressure on the surface element.

With the discrete solution of Equation (5) and Equation (6), we can get the acceleration $\boldsymbol{a}$ and angular acceleration $\boldsymbol{\Dot{\omega}}$ in each time step. The translation velocity and angular velocity are given as:
\begin{equation}
\boldsymbol{v} = \boldsymbol{v^{'}} + \boldsymbol{a} \Delta t 
\end{equation}
\begin{equation}
\boldsymbol{\omega} = \boldsymbol{\omega^{'}} + \boldsymbol{\Dot{\omega}} \Delta t 
\end{equation}
where $\boldsymbol{v^{'}}$ and $\boldsymbol{\omega^{'}}$ are get from previous time step. Then we can get the Eulerian angles around the inertial principal axis and the translation distance in global frame. The solid body coordinate is updated which indicates the solid body motion in this time step is done. Though the orientation of the solid is represented by Eulerian angle $\boldsymbol{\alpha}$ at each time step, the internal calculations on the rotation are based on quaternions to avoid the so called Gimbal lock singularity \cite{Kuipers+1999}. The conversion between them are as follows:
\begin{equation}
\left[\begin{array}{l}
q_0 \\
q_1 \\
q_2 \\
q_3
\end{array}\right]=\left[\begin{array}{l}
\cos \left(\alpha_x / 2\right) \cos \left(\alpha_y / 2\right) \cos \left(\alpha_z / 2\right)+\sin \left(\alpha_x / 2\right) \sin \left(\alpha_y / 2\right) \sin \left(\alpha_z / 2\right) \\
\sin \left(\alpha_x / 2\right) \cos \left(\alpha_y / 2\right) \cos \left(\alpha_z / 2\right)-\cos \left(\alpha_x / 2\right) \sin \left(\alpha_y / 2\right) \sin \left(\alpha_z / 2\right) \\
\cos \left(\alpha_x / 2\right) \sin \left(\alpha_y / 2\right) \cos \left(\alpha_z / 2\right)+\sin \left(\alpha_x / 2\right) \cos \left(\alpha_y / 2\right) \sin \left(\alpha_z / 2\right) \\
\cos \left(\alpha_x / 2\right) \cos \left(\alpha_y / 2\right) \sin \left(\alpha_z / 2\right)-\sin \left(\alpha_x / 2\right) \sin \left(\alpha_y / 2\right) \cos \left(\alpha_z / 2\right)
\end{array}\right],
\end{equation}

\begin{equation}
\left[\begin{array}{c}
\alpha_x \\
\alpha_y \\
\alpha_z
\end{array}\right]=\left[\begin{array}{c}
\operatorname{atan} 2\left[2\left(q_0 q_1+q_2 q_3\right), 1-2\left(q_1^2+q_2^2\right)\right] \\
\operatorname{asin}\left[2\left(q_0 q_2-q_3 q_1\right)\right] \\
\operatorname{atan} 2\left[2\left(q_0 q_3+q_1 q_2\right), 1-2\left(q_2^2+q 3^2\right)\right]
\end{array}\right] .
\end{equation}
Here $\alpha_{i}$ donates teh rotation angle around the i-axis of the instantaneous solid inertia coordinate system and $q$ denotes the set of resulting quaternions. After the solid body motion calculation is done, the translation and rotation velocity treated as the boundary condition are added to the fluid solver.

\subsubsection{Jeffery orbits validation} \label{sec:Validation}

\begin{figure}
\centering
\includegraphics[scale=0.23]{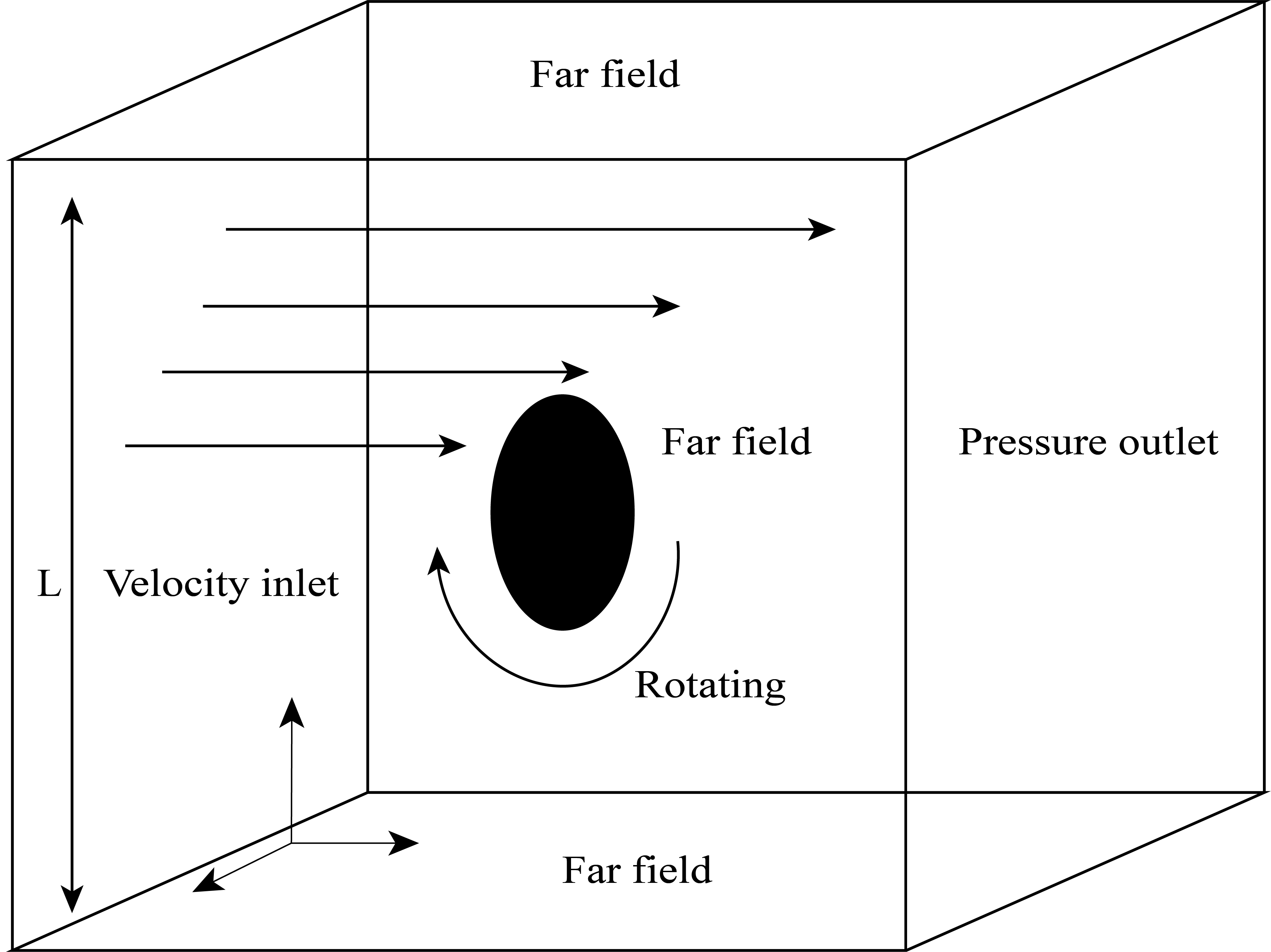}
\caption{Ellipsoid rotation in shear flow.(In order to avoid the boundary influence, a far-field boundary condition is provide)}
\label{SIfig::ellipsoid}
\end{figure}

\begin{figure}
\centering
\includegraphics[width=1\linewidth]{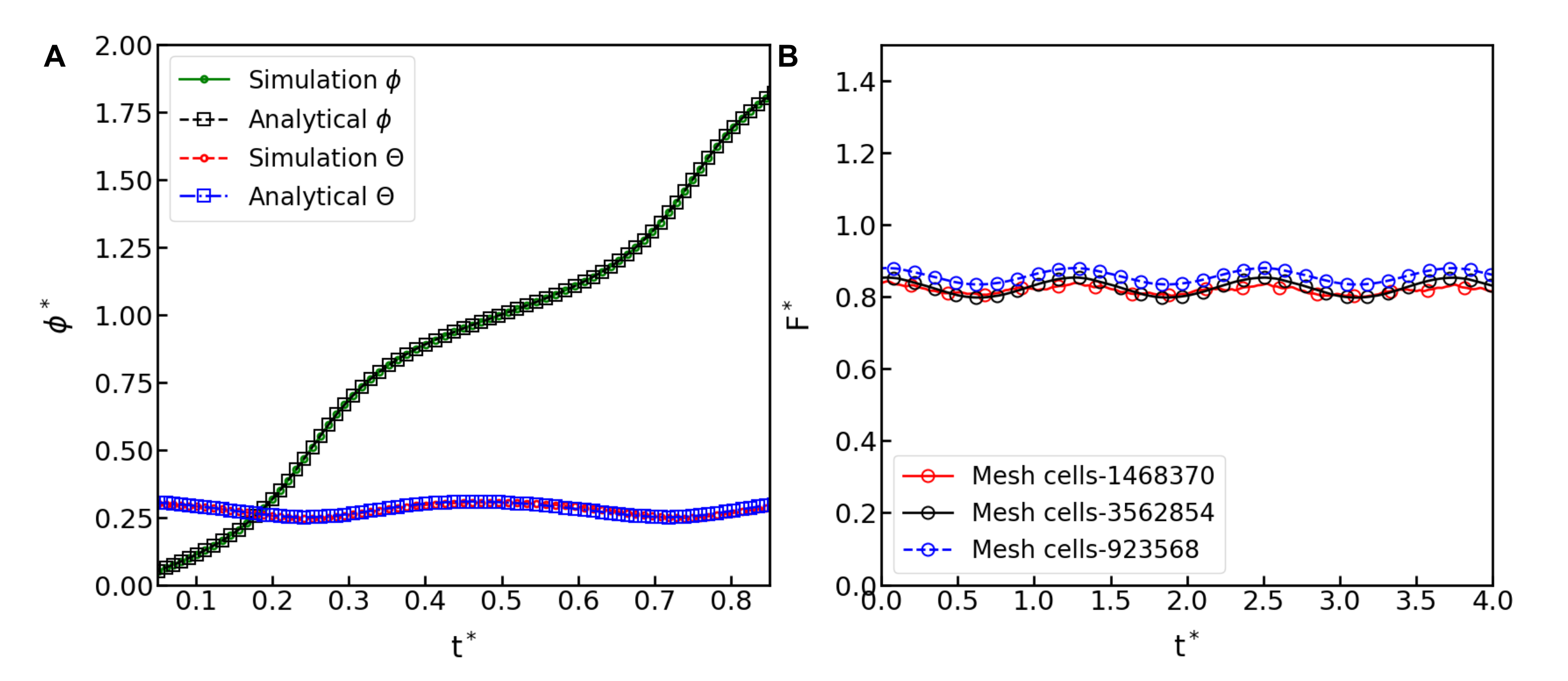}
\caption{(\textbf{A}) Jeffery orbits validation, $\theta$ refers to the angle between the long principal axis of inertia and the vortex direction axis.$\phi$ refers to the angle between the projection of the principal axis of inertia in the xy plane and the long principal axis; (\textbf{B}) Mesh independent.}
\label{SIfig::validationAndMeshIndependent}
\end{figure}

In the Stokes limit of $Re = 0$, Jeffery analytically showed that a single neutrally buoyant ellipsoid in shear flow performs rotary motion. The orbit tracked is the so called Jeffery orbits. At low Reynolds numbers when the effects of inertia are negligible, the solution was confirmed in several experiments by Taylor\cite{Jeffery1922TheMO}.

Here we validate the performance of 6DOF algorithm for classical Jeffery orbit solution. Our set-up contains a neutrally buoyant spheroid subjected to Couette flow at constant shear rate $\Dot{\gamma}$ in a channel. The channel is large enough to eliminate the boundary influence as shown in fig. ~\ref{SIfig::ellipsoid}. The rotation of the solid is then compared against the theoretical result of Jeffery. The analytical Jeffery's angle defined as follows:

\begin{equation}
\tan \theta=\frac{C r}{\left[r^2 \cos ^2 \phi+\sin ^2 \phi\right]^{1 / 2}}, \quad C=\tan \theta_0 \sqrt{\cos ^2 \phi_0+\frac{1}{r^2} \sin ^2 \phi_0}
\end{equation}

\begin{equation}
\tan \phi=-r \tan \left(\frac{\dot{\gamma} t}{r+r^{-1}}\right) .
\end{equation}
 
Here $\theta$ refers to the angle between the long principal axis of inertia and the vortex direction axis.$\phi$ refers to the angle between the projection of the principal axis of inertia in the xy plane and the long principal axis. $r$ refers to the ratio of the ellipsoid's long and short axis, $\theta_0$ and $\phi_0$ are initial rotation angle which set as $\theta_0 = \frac{\pi}{3}, \phi_0 = 0$. fig. ~\ref{SIfig::validationAndMeshIndependent}\textbf{A} shows the rotation of ellipsoid with aspect ratio $r = 2$. The Reynolds number based is $10e^{-2}$. The angle and time shown are scaled by $\pi$ and time value for one period. This result indicates that numerical are in good agreement with theory.

Good agreement between the analytic and numerical in the classical settling and rotation problem indicates that this solver can be used to simulate the flagellum motion in the viscous fluid. Next we will consider the flagellum bundling process numerical and comparing with experiment results.

\subsubsection{Mesh independent}
In fluid mechanics, mesh independent verification ensures the precision and reliability of numerical. When conducting mesh independence verification, selecting appropriate mesh density and refinement strategies is important, such as doubling the number of mesh cells. In this paper, we choose three number of mesh and list the results in fig. ~\ref{SIfig::validationAndMeshIndependent}\textbf{B} (right).

Results show that when mesh number is larger than 1468370. Total force change is small enough. Adding mesh number is a challenge to the computation efficiency. In this paper, we choose middle mesh number.

\subsubsection{Single flagella validation}
\begin{figure}
\centering
\includegraphics[width=0.6\linewidth]{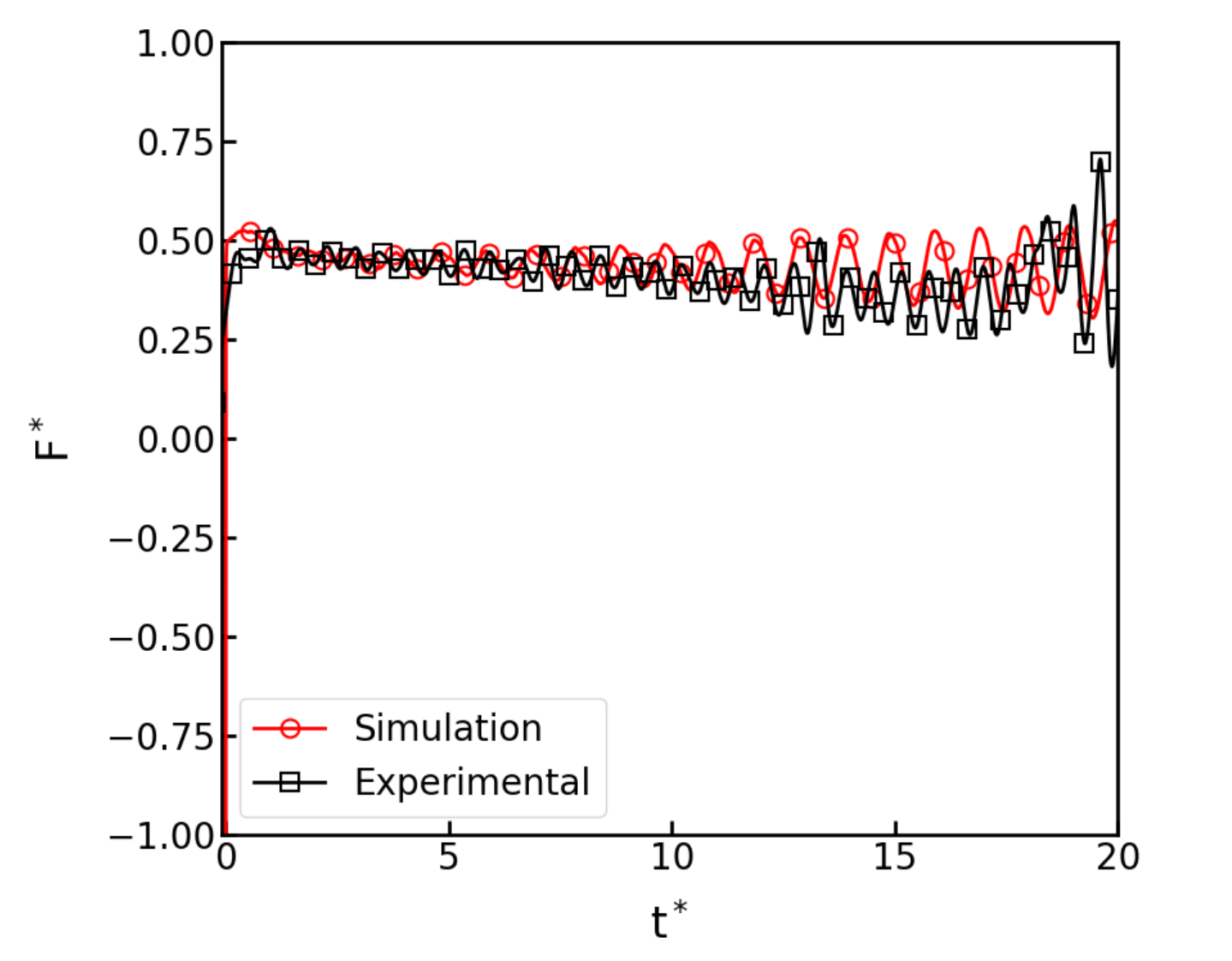}
\caption{Single flagella rotating validation; Red line represents the experiment results and black line represents the numerical results.}
\label{SIfig::single}
\end{figure}

In order to better validate the numerical method, we test single flagella rotating in experiment tank with $\omega = 5.08 rad/s$. Results are shown in fig. ~\ref{SIfig::single}\textbf{B}. The results indicate that the numerical is consistent with the experiment. 

\subsection{Supplementary Text: Mathematical model of bundling process}

\begin{figure}
\centering
\includegraphics[scale=0.25]{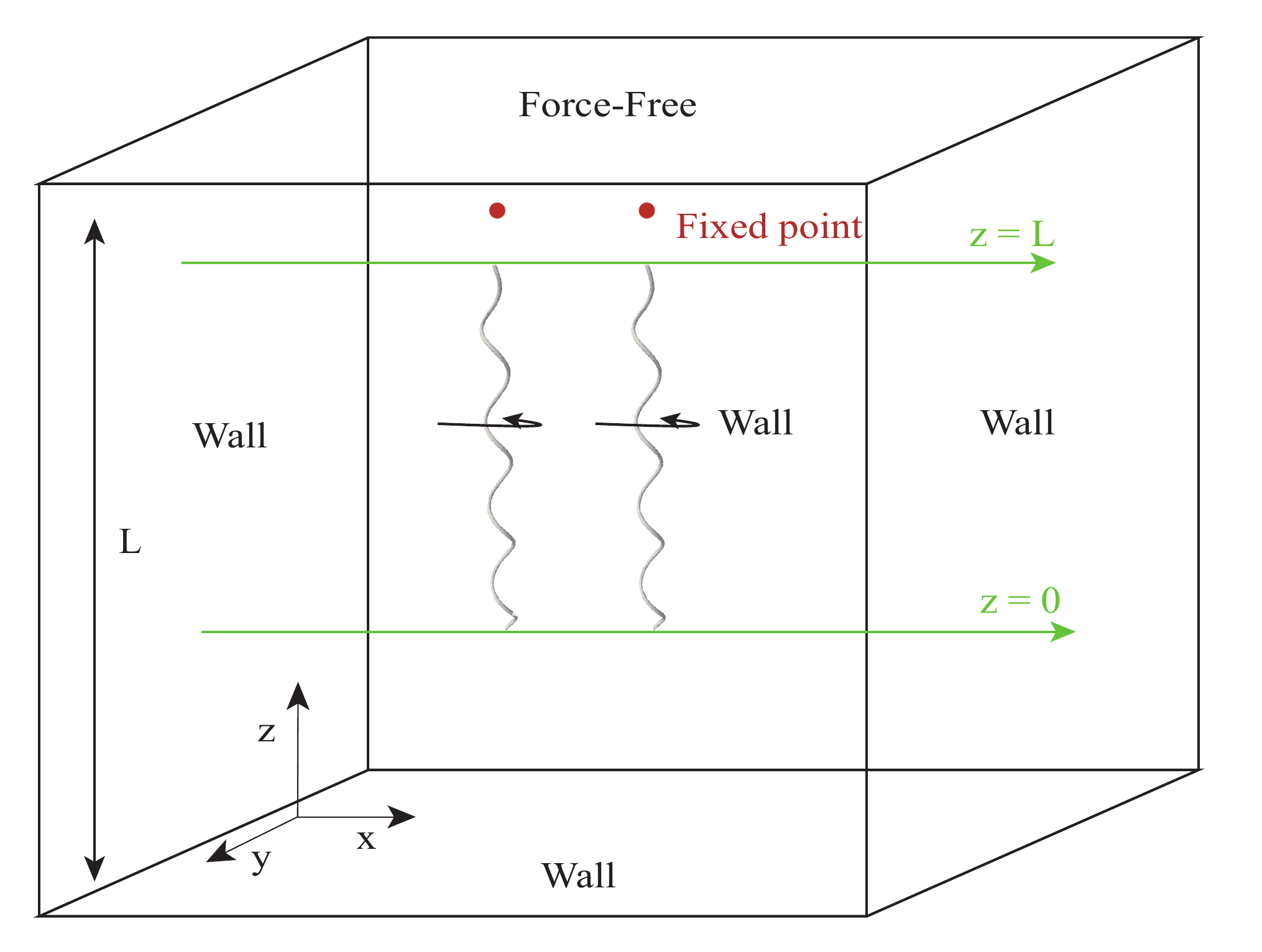}
\caption{Flagella bundling numerical model domain; Faces around flagella and bottom face is set as wall which provide a non-slip boundary condition; Top face is set as a force-free boundary which can not generate external force or backward flow.}
\label{problemdefination}
\end{figure}

In order to match experiment result, the numerical parameters are set same as experiment. The boundary condition are shown in fig. ~\ref{problemdefination}. The boundary faces around flagella and bottom face is set as wall which provide a non-slip boundary condition. In order to avoid the silicone oil-air interface influence, a force-free top boundary which can not generate external force or backward flow is provided.

Parameters of flagellum are same as shown in Table. ~\ref{SITable::helix_parameter}. 

In experiment, two stationary stepper motors under independent control rotate helices in a tank of silicone oil. The viscosity of the silicone oil is $12.5 N\cdot /m3$. The motors typically rotate at $5.08 rad/s$. For these parameters, the Reynolds number $Re \approx 10^{-2}$, low enough to justify neglect of inertial effects. The size of the tank is $300mm \times 300mm \times 300mm $. The helices are made by photosensitive resin.

In numerical, as for the small deformation of photosensitive resin in the experiment process, a rigid body model is introduced. To model the experiment motor rotating with constant angular velocity. A simple mathematical model are introduced. Two flagellum are actively rotating with a constant angular velocity around two fixed points. Under this assumption, the Newton’s second law is not needed. Considering the rotation axis is moving during bundling process. To get constant angular velocity on rotation axis, by using Hooke’s law, an additional spring torque is added to Equation (6), the final Euler equation is listed as follows:
\begin{equation}
\mathcal{I}_s \frac{\boldsymbol{d} \boldsymbol{\Omega}}{\boldsymbol{d} t}= \boldsymbol{M}_O - K \cdot (\boldsymbol{\Omega_{0}}-\boldsymbol{\Omega}).
\end{equation}
Here $\boldsymbol{\Omega_{0}}$ is the constant angular velocity around rotation axis which is set to be same as experiment, $K$ is the stiffness. If $K$ is set large enough, the solved $\boldsymbol{\Omega}$ on the rotation axis is infinitely close to the defined angular velocity $\boldsymbol{\Omega}_0$. This method is widely used in numerical numericals \cite{SHU20071607}. By discrete solving Equation (18), we can simulate the bundle process.

\subsection{Hydrodynamics force along x direction}
\begin{figure} 
    \centering
    \includegraphics[width=1\linewidth]{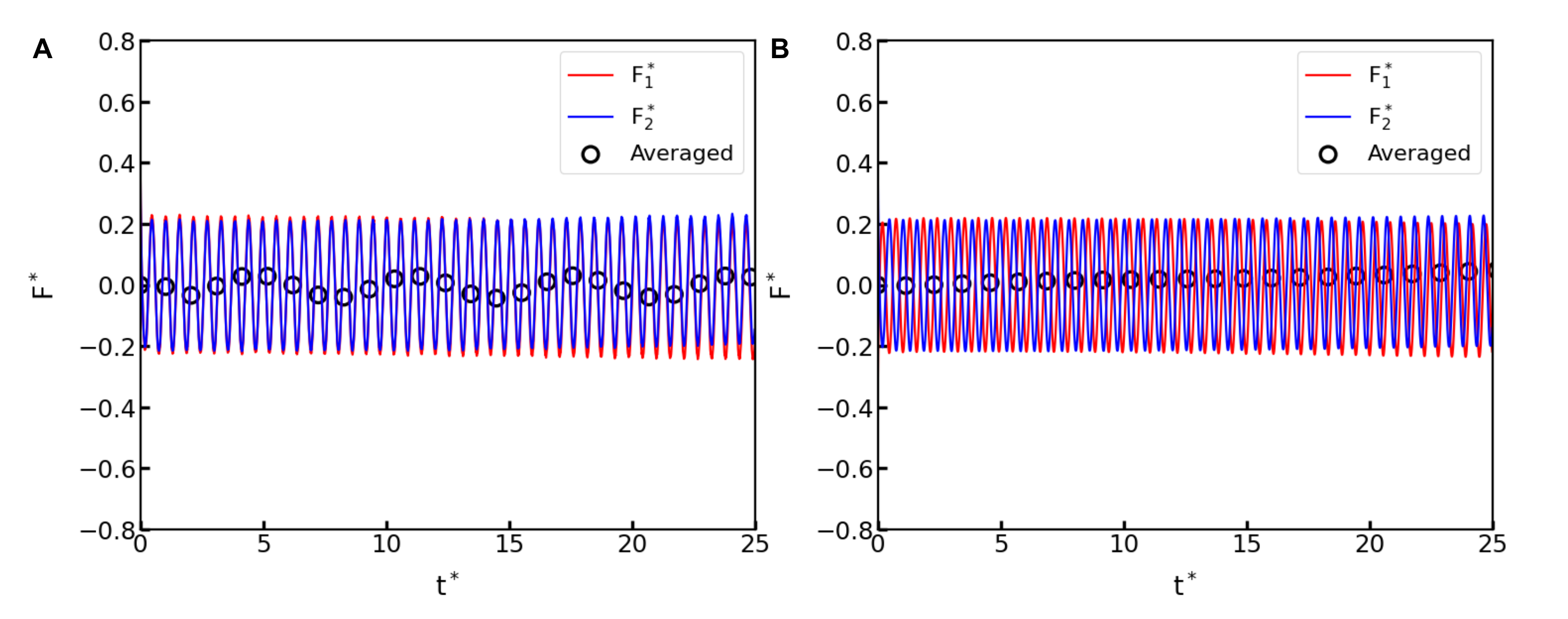}
    \caption{(\textbf{A}) Pure hydrodynamics force acting on the system along the x direction with $\Delta \Phi = 0$;(\textbf{B}) Pure hydrodynamics force acting on the system along the x direction with $\Delta \Phi = \pi$.(Red line and blue line represent single flagella hydrodynamic force along x-direction;Circle symbol represents pure hydrodynamic force of two flagella along x-direction).}
    \label{SIfig:XforceCompare}
\end{figure}

fig. ~\ref{SIfig:XforceCompare} indicates that the helices lean towards one of them at the steady state is a pure hydrodynamics force acting on the system along the $x$ direction.fig. ~\ref{SIfig:XforceCompare}\textbf{A} gives the pure hydrodynamics force acting on the system along the x direction with $\Delta \Phi = 0$. fig. ~\ref{SIfig:XforceCompare}\textbf{B} gives the pure hydrodynamics force acting on the system along the x direction with $\Delta \Phi = \pi$.

\subsection{Correlation coefficient of fluid velocity field at different times}
\begin{figure} 
    \centering
    \includegraphics[width=0.6\linewidth]{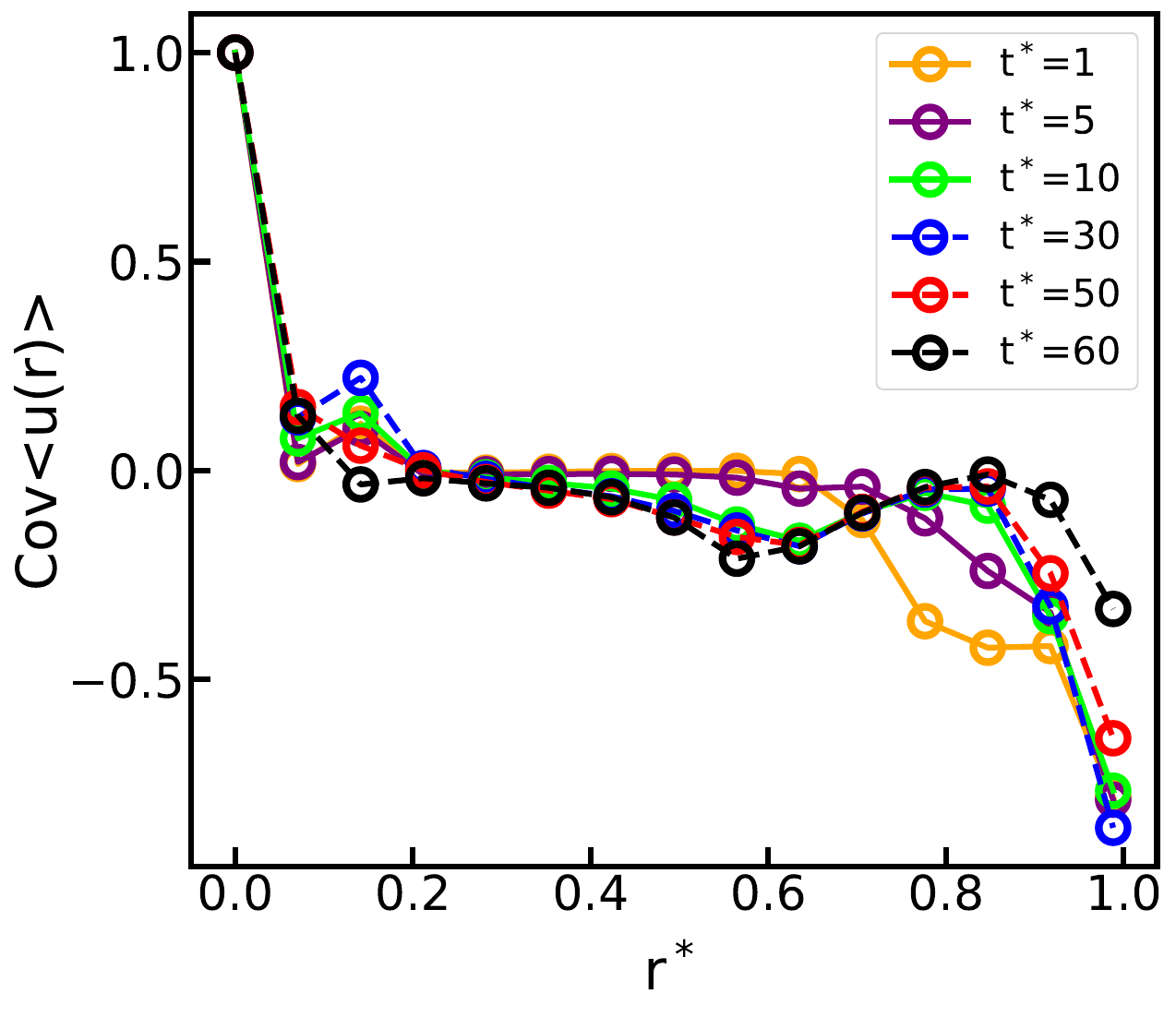}
    \caption{Correlation coefficient at different times in bundling process (xy-cut plane). $r^*$ is normalized by $r/r_{max}$. $r_{max}$ is the maximum length between two arbitrary points in the flow filed. $t^*$ is the instantaneous time.}
    \label{SIfig:coefficient}
\end{figure}

In order to get the fluid field structure variation during the bundling process, we introduce the correlation coefficient as follows:

\begin{equation}
Cov<\boldsymbol{U}(\boldsymbol{r})>=\frac{\langle \boldsymbol{U}(\boldsymbol{r}) \cdot \boldsymbol{U}(0)\rangle}{\left\langle|\boldsymbol{U}(0)|^2\right\rangle}
\end{equation}

fig. ~\ref{SIfig:coefficient} indicates that in the beginning of the bundling process, the fluid field changes large, when the saddle structure is formed (see Fig. 4B), the correlation coefficient basically unchanged. Correlation coefficient of $0.2 < r^* < 0.7$ becomes larger as time grows. This tendency indicates the scope of bundling influence is gradually expanding.

\subsection{Supplementary Text: Model illustration}
\begin{figure}[htbp] 
    \centering
    \includegraphics[width=0.3\linewidth]{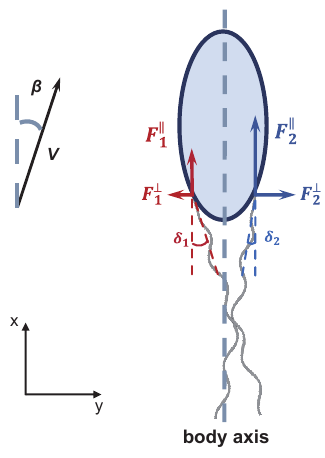}
    \caption{Illustration for the simplified model in the main text.}
    \label{SIfig::Theomodel}
\end{figure}

As fig. ~\ref{SIfig::Theomodel} shows, the theoretical model is presented here. 

\clearpage
\bibliographystyle{science}  
\bibliography{scibib}  
